\pdfoutput=1
\documentclass[preprint]{emulateapj}
\usepackage{rotating}
\usepackage[version=3]{mhchem}
\usepackage[inline]{enumitem}
\usepackage{color}
\RequirePackage{xspace}
\usepackage{multirow}

\def\ifundefined#1{\expandafter\ifx\csname#1\endcsname\relax}

\ifundefined{ensuremath}\def\ensuremath#1{\relax\ifmmode{#1}}
\else${#1}$\fi\else\relax\fi
\ifundefined{nuc}\def\nuc#1#2{\relax\ifmmode{}^{#1}{\protect\mathrm{#2}}
\else${}^{#1}$#2\fi}\else\relax\fi

\newcommand{\unit}[1]{\ensuremath{\,\, \mathrm{#1}}}
\newcommand{\mch}{M_{\rm Ch}}

\shorttitle{Type Ia Supernovae: Taking the Magic out of CMAGIC, and
other Flaws of Nature}

\shortauthors{Hoeflich et al.}

\begin{document}

\title{Light and Color Curve Properties of Type Ia Supernovae: Theory vs. Observations}   

\author{P. {Hoeflich}\altaffilmark{1*}} 
\author{E.~Y. {Hsiao}\altaffilmark{1}}
\author{C.~ {Ashall}\altaffilmark{2}} 
\author{C.~R. {Burns}\altaffilmark{3}} 
\author{T.~R. {Diamond}\altaffilmark{4}}
\author{M.~M. {Phillips}\altaffilmark{5}}
\author{D. {Sand}\altaffilmark{6}} 
\author{M. D. {Stritzinger}\altaffilmark{7}}
\author{N. {Suntzeff}\altaffilmark{8}} 
\author{C. {Contreras}\altaffilmark{9}} 
\author{K. {Krisciunas}\altaffilmark{8}}
\author{N. {Morrell}\altaffilmark{9}}
\author{L. {Wang}\altaffilmark{8}}

\altaffiltext{1}{Department of Physics, Florida State University, Tallahassee, FL 32306, USA, * \url{phoeflich77@gmail.com}}
\altaffiltext{2}{Astrophysics Research Institue, Liverpool John Moore University, 146 Brownlow Hill, Liverpool L3 5RF, UK}
\altaffiltext{3}{Observatories of the Carnegie Institution for Science, 813, Santa Barbara St., Pasadena, CA 91101, USA}
\altaffiltext{4}{NASA Goddard Space Flight Center, Greenbelt, MD 20771, USA}
\altaffiltext{5}{Carnegie Observatories, Las Campanas Observatory, Casilla 601 La Serena, Chile}
\altaffiltext{6}{Physics \& Astronomy Department, Texas Tech University, Box 41051, Lubbock, TX 79409-1051, USA}
\altaffiltext{7}{Department of Physics and Astronomy, Aarhus University, Ny Munke Gade 120, 8000, Aarhus, Denmark}
\altaffiltext{8}{The G.P. \& C. Woods Mitchell Institute for Fundamental Physics \&  Astronomy, Texas A\& M University, 
Dept. of Physics \& Astronomy, 4242 TAMU, College Station, TX 77843, USA}
\altaffiltext{9}{Departamento de Ficica, Universidad Tecnicia Federico Santa Maria, Ava Espaa 1680, Casilla 110-V, Valparaiso, Chile}

\begin{abstract}
We study optical light curve(LC) relations of type Ia supernovae(SNe~Ia)
for their use in cosmology using high-quality photometry published by the {\em Carnegie-Supernovae-Project}(CSP-I).
We revisit the classical luminosity-decline-rate ($\Delta m_{15}$) relation and the Lira-relation, as well as investigate the time
evolution of the ($B-V$) color { and $B(B-V)$, which serves as the basis of the color-stretch relation and 
 Color-MAGnitude-Intercept-Calibrations(CMAGIC). 
Our analysis is based on explosion and radiation transport simulations for spherically-symmetric
delayed-detonation models(DDT) producing normal-bright and 
subluminous SNe~Ia. Empirical LC-relations can be understood as having the same physical underpinnings: i.e. the opacities,
ionization balances in the photosphere, and radioactive energy deposition changing with time from below to above the photosphere.
 Some $3-4\unit{weeks}$ past maximum, the photosphere recedes to \ce{{}^{56}Ni}-rich layers 
 of similar density structure, leading to a similar color evolution. 
An important secondary parameter is the central density $\rho_c$ of the WD because at higher densities more electron capture elements 
are produced at the expense of \ce{{}^{56}Ni} production. This results in a $\Delta m_{15}$ spread of $0.1\unit{mag}$ for 
normal-bright and $0.7\unit{mag}$ in sub-luminous SNe~Ia and $\approx 0.2\unit{mag}$ in the Lira-relation.
We show why  color-magnitude diagrams emphasize the transition between physical regimes, and enables the construction
of  templates which depend mostly on $\Delta m_{15}$ with little dispersion in both the CSP-I sample and our  
DDT-models. This allows to separate intrinsic SN~Ia variations from the interstellar reddening characterized by $E(B-V)$ and $R_{B}$. 
Invoking different scenarios causes a wide spread in empirical relations which may suggest one dominant scenario.}
\end{abstract}


\section{Introduction}\label{Introduction}
\label{intro}

Much of the progress in the use of Type Ia supernovae (SNe~Ia) as extragalactic distance indicators has resulted from the discoveries of 
empirical relations describing  the evolution of  their light curves (LC) and color curves  as a function of their luminosity.  
These empirical relations highlight the incredible uniformity for the vast majority of SNe~Ia. 
\citet{p93} established the correlation between the absolute $V$-band peak brightness, $M_V$, of a SN~Ia and its decline in brightness fifteen days later,  
parameterized by $\Delta m_{15}$.\footnote{{ The light curve decline rate parameter, $\Delta m_{15}$, is found to be correlated to the peak luminosity  of a SN~Ia in the sense that more luminous objects 
exhibit slower decline rates \citep{p93}.}}
Using $\Delta m_{15}$ to calibrate SNe~Ia dramatically lowers the Hubble diagram dispersion. 
More evidence for the apparent homogeneity of the population comes from the observations of fairly uniform color evolution for at least normal-bright SNe~Ia. 
\citet{Riess96} and \citet{Tripp98} introduced another calibration reflecting that fainter SNe~Ia tend to be redder { around maximum light}.
\citet{lira95} and \citet{phillips99} found that the intrinsic $(B-V)$ colors are nearly identical between $30-60\unit{days}$ after 
$B$-band maximum light (hereafter $T(B)_{max}$), despite a wide range in $\Delta m_{15}$, and \citet{2003AJ....125..166K} showed similarities in the evolution of 
optical and infrared (IR) colors for normal-bright SNe Ia. 
Moreover, most SN~Ia light curves evolve along the same path in a color-magnitude diagram, which inspired the {
Color-MAGnitude Intercept Calibration (CMAGIC) method for the distance determination of SNe~Ia \citep{wang2002}.\footnote{{ All CMAGIC methods are based on the empirical finding that SNe~Ia follow a tight path in absolute magnitude-color space with a fast change in slope around maximum light. Early approaches use both the rise and the 
decline and multiple colors, but most of the recent implementations  
use $B$ and $V$ and the linear slope past maximum light \citep{2006ApJ...641...50W,2006ApJ...644....1C}. In the following, we will
refer to CMAGIC as the method based on the characteristic shape of the brightness color relations in $B$ and $V.$}}}
Most recently, \citet{burns11} used a correlation between the timing of the reddest $(B-V)$ color (parameterized by the color stretch parameter $s_{BV}$ \footnote{$s_{BV}$  is a dimensionless quantity and is  defined as the epoch of maximum $(B-V)$ divided by 30 days.}) and maximum brightness to 
obtain even  a better standardization than $\Delta m_{15}$ for fast-declining SNe~Ia. 

With the advent of high-quality data, we are beginning to see diversity of SNe~Ia upon more detailed analysis. 
The physics behind the $\Delta m_{15}$ relation may be understood as a result of { rapid and steep drops of the
opacities with temperature, namely their flux and Rosseland means. This has been shown  for several classes of explosion models over the last 20 years}
  \citep{HKWPSH96,nughydro97,umeda99,2000ApJ...530..744P,HGFS99by02,kasen09}. However,
the underlying physics of these empirical color relations has not been well explained.

The goal of this paper is to provide a basic understanding of the empirical $(B-V)(t)$ relation, the Lira relation, {the brightness color
diagrams with their characteristic shapes and, thus,  the origin of color-stretch 
standardizations, CMAGIC.}  
In addition, we examine the robustness of these empirical relations against varying filter functions, reddening, explosion conditions, and progenitor properties.
This study is not only important for a better understanding of the SNe~Ia explosion physics but also in their use as tools for 
high-precision cosmology.  A better understanding may also help to improve the accuracy of existing empirical methods and in developing new ones. 
{ We emphasize that no individual SNe~Ia are analyzed or fitted, but rather, they are used as tracers of the relations. 
This allows us to use models and theoretical results previously published, in parts, 
many years before high precision photometry became available or prior to the discovery of today's known color relations. This avoids adjusting model parameters a posteriori. The models 
are used as tracers of the relations, but the loci of models and individual SNe~Ia
do not coincide.  For high precision, we use broad-band LCs (energy distribution) because of their 
stability with respect to small variations in conditions at the photosphere. As shown by 
\citet{1983ApJ...270..123B}, to first order, spectra can be understood as overlapping P-Cygni 
profiles with Doppler shifts less than the typical width of broad-band filters. In contrast to LCs, 
spectra measure the abundances and velocities at the photosphere 
and the Doppler shifted lines. Because nuclear physics determines the composition and specific energy release of  C-O WDs,
most scenarios for thermonuclear explosions will show a generic evolution
from spectra dominated by unburned material to elements of explosive C-burning (O/Ne/Mg) and incomplete burning (Si/S)
to iron group elements \citep[see][and references therein]{2006NewAR..50..470H}. However, spectral details are sensitive
to small variations at the photosphere and fits to individual SNe~Ia require numerically expensive iterative tuning 
of parameters related to the progenitors and simulations of explosion models
or parameterization of the photospheres for a given time. Here, we study relations without fitting individual SNe~Ia, and a few spectral time-series are shown only for demonstration purposes.} 

We compare theoretical light curves with observations from the {\em Carnegie Supernovae Project} (CSP-I; \citealt{2006PASP..118....2H, contreras2010, 2010AJ....139..120F, 2011AJ....142..156S}) to evaluate the validity of the models and their limitations. 
To avoid posterior tuning of model parameters to {\it match} observations, we use {\it existing} explosion models, theoretical LCs 
{\it previously} published by our group without selecting models based on individual observations { and published 
long before high-precision multi-band LCs became available for a large set of SNe~Ia}.
To cover the parameter-space of SNe~Ia, including subluminous objects, we have included new simulations using identical methods, 
discretization, and input physics. { We reference several of our spectral studies for the optical to mid-infrared wavelength regions 
and show some examples for the evolution of optical spectra.}
 For the direct comparison with the CSP-I observations, the monochromatic fluxes
have been recalculated for all simulations using the CSP passbands.  
{ 
Our study is possible only now because of the large sample of high-precision multi-band LCs observed with a single stable and
well-characterized photometric system by CSP.}

From the theoretical perspective a wide variety of explosion scenarios  can be expected to produce SNe~Ia.
The observed diversity in LCs and spectra (e.g., the evolution of Doppler shifts of lines) may suggest multiple progenitor channels and 
explosion scenarios.  
Therefore, we discuss our results obtained with our baseline models in the context of other scenarios.
This study is based on a set of classical Chandrasekhar mass ($\mch$) carbon-oxygen (C/O) white dwarfs (WDs)  spherical delayed-detonation 
models (DDT) \citep{khokhlov91,yamaoka92,gamezo04,p11}. 
The suite of models involve a transition from deflagration to detonation burning parameterized by the transition density, $\rho_{\rm tr}$.
While a wide variety of explosion mechanisms may be possible in nature, we choose the DDT scenario because it predicts 
some of the basic characteristics of SNe~Ia:  overall spherical density distributions as indicated by small continuum polarization 
\citep{howell01,mound10a,patat12}, overall spherical geometry of SN~Ia remnants (SNR) with layered
chemical structures \citep{rest05,2006ApJ...645.1373B,fesen07,rest08}, ``broad'' emission line profiles and stable \ce{^{58}Ni} at late times indicating electron 
capture in central regions \citep{Gerardy,h04,motohara06,maedanature10, 2015A&A...573A...2S,tiara15}, and evidence for a deflagration phase in 
S-Andromeda \citep{fesen16,2016AAS...22723811W}. 
We use spherical models because they suppress mixing during the deflagration phase, which is required to reproduce observed IR 
spectra signatures and the tightness of the luminosity vs. $\Delta m_{15}$ relation.

Within the class of DDT models, $\Delta m_{15}$ is well-understood since LCs are powered by the radioactive decay of \ce{{}^{56}Ni} \citep{ColgateMckee69}.  
More \ce{{}^{56}Ni} causes the envelopes to be hotter, increasing the luminosity of the SN. 
A higher temperature means higher opacities and, thus, longer diffusion time scales and ultimately a slower decline rate after maximum light 
\citep{HKWPSH96,nughydro97,umeda99,kasen09}. 
Although the tightness of the $\Delta m_{15}$ relation is understood within the framework of spherical models 
\citep{HKWPSH96,HGFS99by02,2010ApJ...710..444H}, it falls apart when taking into account instabilities produced in 3D deflagration burning models
\citep{HKWPSH96,kasen09,baron} as predicted \citep{khokhlov95,n95,livne99,rein99,gamezo03,roepke06,plewa2007,p11}.
High magnetic fields have been suggested to suppress mixing induced by Rayleigh Taylor instabilities (RT) and this idea is consistent with some observations 
\citep{h04,penney14,Remming14,tiara15,boyan16}.

The organization of this paper is as follows.
In Sect.~\ref{Observations} we briefly describe the CSP-I data set used in this study, while 
in Sect.~\ref{Methods} our theoretical optical LCs and color curves are presented, as well as the empirical relations, and we then discuss their physical origins 
and expected variations within the DDT scenario. 
In Sect.~\ref{Analysis} we confront theory with a selected set of observed SNe~Ia, including both normal-bright and subluminous objects, in order to establish agreement on a qualitative level without fine-tuning of the models. 
Observable properties, redshift distances, interstellar medium (ISM) reddening correction, and the time of maximum brightness are taken 
from empirical fits. 
In Sect.~\ref{Reddening} we will develop a physics-based approach using $\Delta m_{15}$, $B(B-V)$ and $V(B-V)$ and theoretical templates in both 
$B$- and $V$-band in order to optimize fits of light curves, color curves, and new empirical relations. 
This allows for a detailed comparison of the theoretical and observed shapes of empirical relations. 
The fit parameters of individual LCs can be directly interpreted as distances, ISM reddening, explosion times, and model properties. 
In Sect.~\ref{Analysis2} we present some checks for the consistency of theoretical templates with the observed sample and confront our 
findings for distance and reddening with the corresponding values based on empirical methods. 
We discuss how the consistency checks may be used to verify properties and for error estimates, in addition to how theoretical 
parameters may lead to improved fully empirical methods which may be used for high-precision SN cosmology.
Finally, in Sect.~\ref{Alternatives} our results are placed into the context of other explosion scenarios to evaluate whether the underlying physics 
driving their predicted observational properties are generic to thermonuclear explosions in general.

%
%

\section{Observations}{\label{Observations}}
%
%

In this paper, we use $B$- and $V$-band LCs of local SNe~Ia, located within a redshift range of $z = 0.004-0.034$. 
The photometry considered here is contained within the  final CSP-I SNe~Ia data release (Krisciunas et al., in preparation), which was obtained over five 1-year observational 
campaigns using facilities located at the Las Campanas Observatory. 
The photometry is in the {\em natural system} of the CSP-I. 
Choosing to work with only CSP-I data means dealing with a single well-understood and stable photometric system, resulting in 
well-calibrated data within a typical uncertainty of $0.01\unit{mag}$. 
$K$-corrections were performed using spectral templates constructed by \citet{2007ApJ...663.1187H} and implemented within the light curve fitting software package  {\tt SNooPy} \citep{burns11}.

The final CSP-I data release consists of 123 { SNe thought to have a white dwarf origin.}
In the discussion of the brightness-decline relation, forty-three SNe~Ia with sufficient phase coverage from pre-maximum light to the first 
inflection point in the LCs were used. 
To facilitate the detailed comparisons with model LCs and color curves, twenty SNe~Ia from CSP-I with adequate phase coverage from 
pre-maximum to $50\unit{days}$ past maximum are used. 
In addition we also examine the SNe~Ia pair, SNe 2007on and 2011iv, that reside in the same host galaxy for a case study of distance 
determination. 
The list of these SNe~Ia, their host galaxies, redshifts corrected for the cosmic microwave background (CMB) dipole, and decline rates 
$\Delta m_{15,s}$  are tabulated in Table~\ref{table_obs}. { The parameter $\Delta m_{15,s}$ is defined as the decline in brightness 
over $15\unit{days}$ measured in the stretched time $s \times t$ with $s$ as the stretch parameter from \citet{goldhetal01} (see Sect. \ref{LC}).
These quantities are model-independent and are used as a baseline for comparison throughout this paper.}

\section{Theory: on the origin of empirical relations}{\label{Methods}}
\subsection{Background models}\label{bgm}

\begin{figure}   
\includegraphics[angle=360,width=0.9\textwidth]{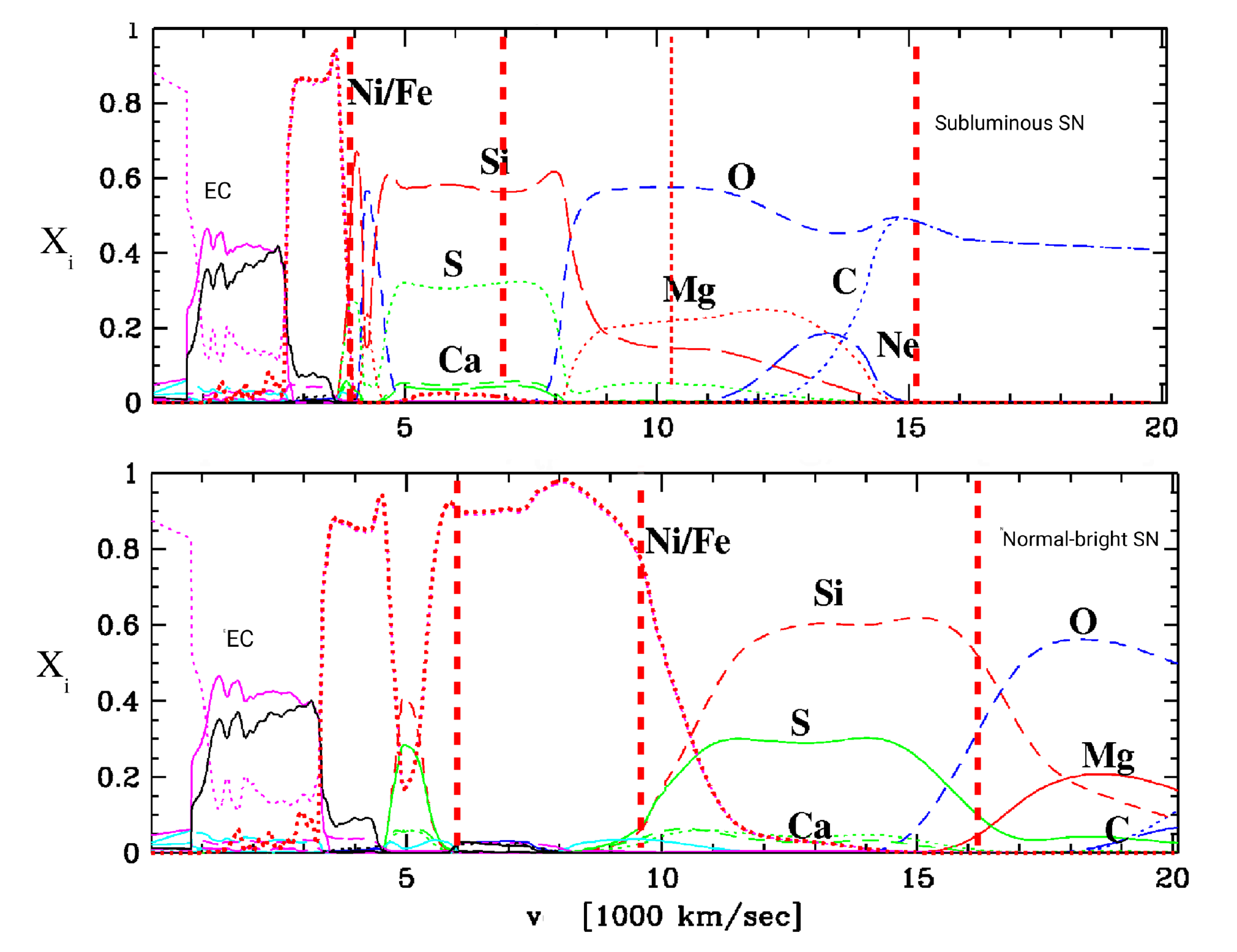} \\
\caption{
Typical chemical structures of classical, spherical DDT models with solar metallicity for a normal-bright (Model 23, the 
reference model, bottom) and a subluminous, SN~1991bg-like, object (Model 08, top) from \citet{HGFS99by02}. 
We give the mass fractions $X_i$ as a function of the expansion velocity, $v$.
The central abundances are dominated by stable isotopes of iron-group elements. 
In addition, the vertical bars indicate the position of the {\sl Thompson-scattering} photosphere  in velocity space at $10$, $20$, and $30\unit{days}$ (right to left, 
thick dotted line) after the explosion. 
For the subluminous model (top), the location of the photosphere is indicated at $2.5\unit{days}$ by the thin dotted line. 
Note that, soon after the explosion, the photospheric velocity increases for several hours for normal-bright SNe and up to three days in 
SN~1999y-like models \citep{2006NewAR..50..470H}.
} 
\label{models}
\end{figure}

\begin{figure}   
\includegraphics[angle=360,width=0.5\textwidth]{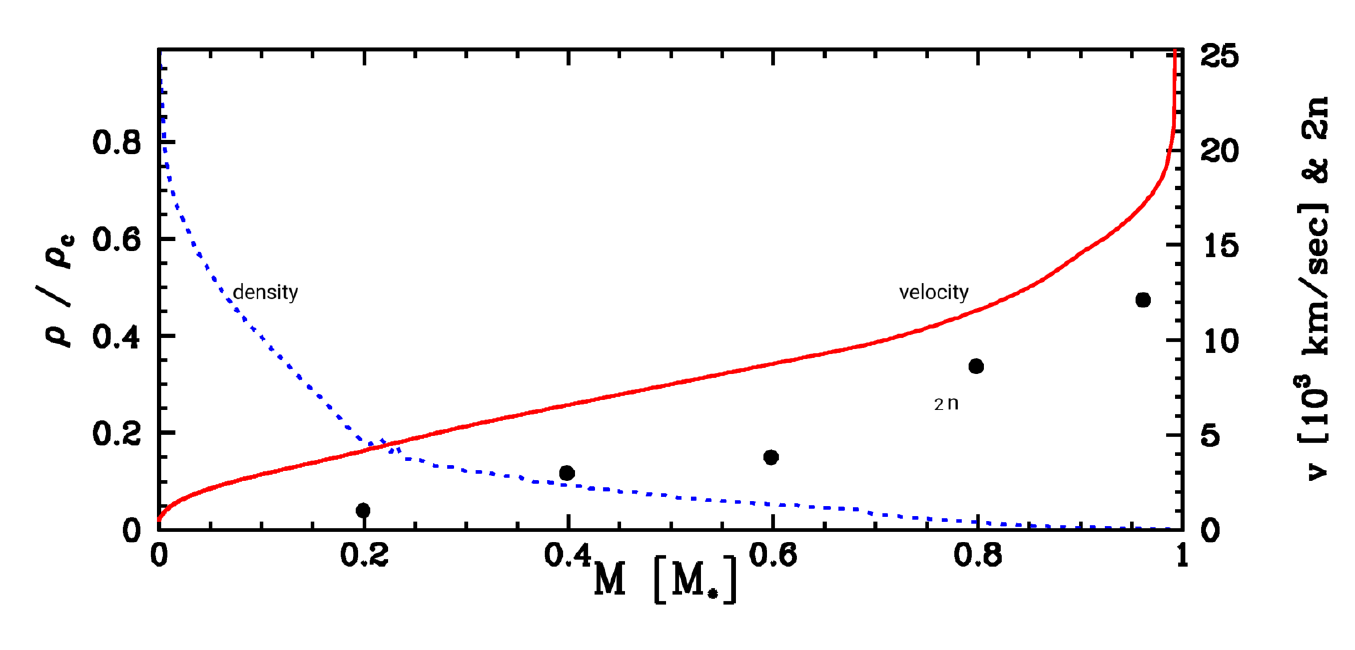} 
\includegraphics[angle=360,width=0.45\textwidth]{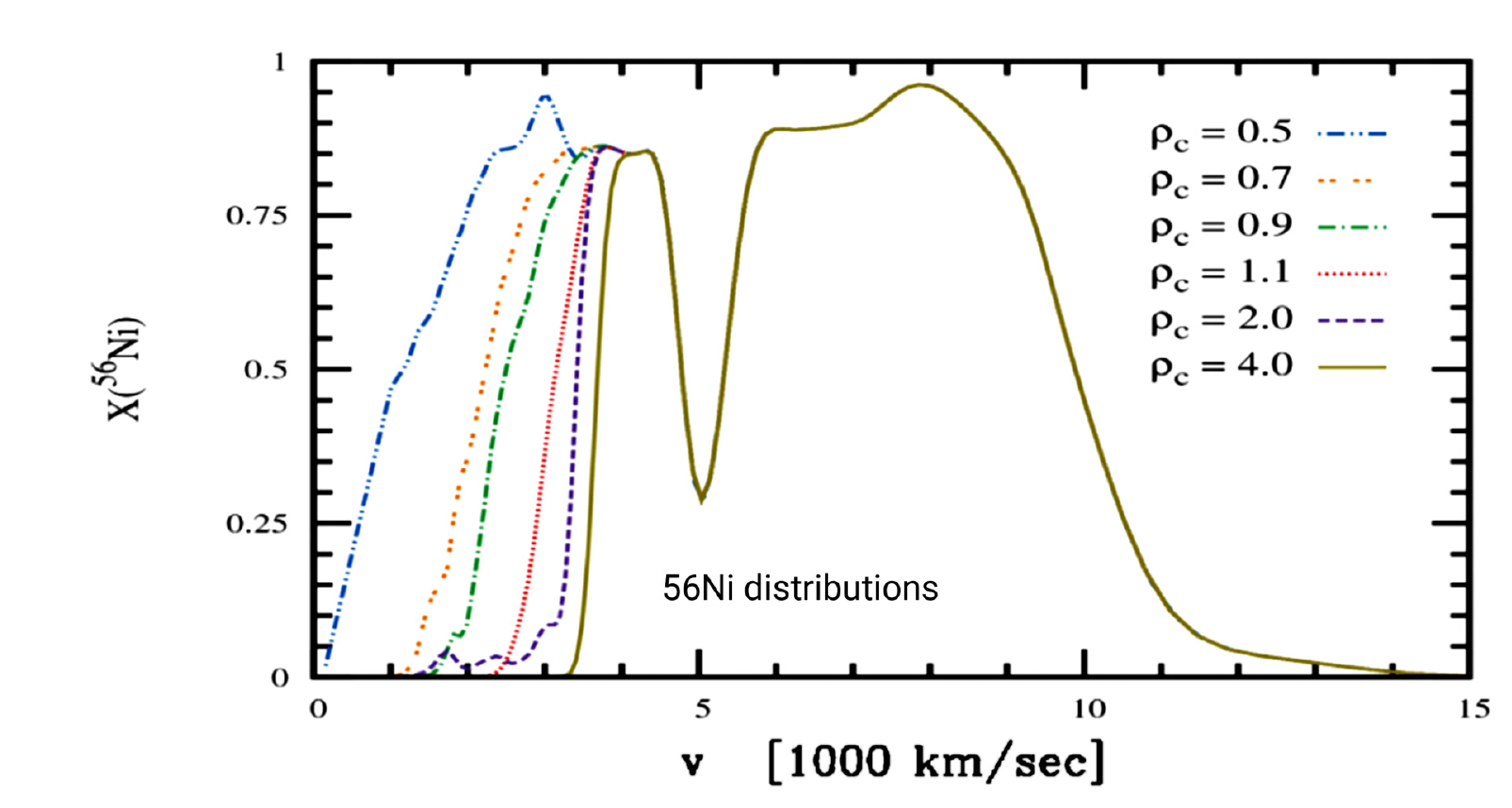}
\caption{
Models with $\rho_{\rm tr}$ of $2.3\times10^7\unit{g}\unit{cm^{-3}}$. 
On the left, the velocity (red line, right scale), normalized density (blue dotted, left scale), and density gradient $2\times n$ with 
$\rho \propto r^{-n/2}$ (black dots, right scale) are given for the normal-bright Model 23.  
$n$ has been averaged over $0.1\unit{M_{\rm WD}}$.  
In the plot on the right, we show the influence of the initial central density $\rho_c$ (in $10^9\unit{g}\unit{cm^{-3}}$) on the \ce{{}^{56}Ni} 
distribution. 
The transition density has been varied slightly, $0 ... 3 \%$,  to force very similar outer layers. 
Note that the density structures and central regions are very similar over a wide range of DDT models \citep{HGFS99by02,tiara15}.
} 
\label{models_rho}
\end{figure}

For this study we use a spherical DDT model in which the explosion occurs in a $\mch$ C/O WD. 
After an initial deflagration phase, a detonation is triggered when the density at the flame front drops to a transition density 
$\rho_{\rm tr}$, which is a parameterization of the amount of burning during the deflagration. 
The progenitor introduces a number of additional free parameters: 
\begin{enumerate*}
\item[a)] the primordial metallicity, $Z$, of the progenitor star;
\item[b)] the ratio of carbon to oxygen as a function of mass coordinate, $m$, in the WD, which mostly depends on the total mass of the 
progenitor and, to a lesser extent, on $Z$ \citep{dominguez02}, and;
\item[c)] the central density, $\rho_c$, of the WD at the time of the explosion.
\end{enumerate*}
The basic parameters and some of the observable properties as measured from the model LCs are listed in Table~\ref{table_mod}, { and
some typical structures are given in Figs. \ref{models} and \ref{models_rho}}.
 In this paper, we will refer to specific models by the names given in Column 1 of Table~\ref{table_mod}.
{ The simulations are based on our HYDrodynamical RAdiation code (HYDRA), which utilizes a nuclear network of 218 isotopes during the early phases of the explosion and detailed, time-dependent non-LTE 
models for atomic level populations, including $\gamma$- and positron transport and radiation-hydrodynamics to calculate  
LCs and spectra \citep{h95,hoeflich2003hydra,penney14}}.\footnote{ We include hydrodynamics because the energy production by radioactive \ce{{}^{56}Ni} decay 
is equivalent to a velocity of $\approx 3{,}000\unit{km}\unit{s^{-1}}$.
 The effect of \ce{{}^{56}Ni} decays may have been observed in the supernova remnant S-Andromeda that exhibits
caustic structures in the \ce{Fe} distribution, which can be understood as the imprint of ``under-developed'' Rayleigh-Taylor instabilities 
in a mixture of stable and radioactive elements of the iron group
\citep{fesen07,fesen15}.} { We note that positron-transport is not important for early LCs, however it is critical for several of the
spectral tests of our models cited below.}
For the LCs we solve the radiation transport equations in co-moving frames using $\approx100$ frequency groups.
For consistency between new and previously-published models, all simulations use the same atomic models with line transitions taken 
from the compilations of \citet{kurucz93}, supplemented by forbidden line transitions of iron group elements in the lower ionization states 
\citep{2015ApJ...798...93T}.  
During the deflagration phase, the rate of burning is parameterized based on physical flame models and calibrated to three-dimensional 
(3D) hydrodynamical models \citep{2000ApJ...528..854D}.

These models have been widely applied to analyze { and fit} individual SNe~Ia, particularly flux in the optical and IR and polarization 
spectra \citep{h95,hwt98,1998ApJ...496..908W,lentz01,quimby06,baron,patat12,2015MNRAS.454.2549B,tiara15,2015ApJ...798...93T}.
Note that very similar chemical abundances and velocity structures have been obtained by the analysis of observed spectral sequences 
using the method of ``spectral tomography''.
This method uses time-series observations of SNe~Ia to reconstruct abundance structures rather than producing them using explosion 
models \citep{2008MNRAS.386.1897M,2009MNRAS.399.1238H,2011MNRAS.410.1725T,2014MNRAS.445.4427A}. 

Throughout this paper we use the Swope filters \citep[see][]{2011AJ....142..156S} to compare our theoretical results with the CSP-I light curves.
For the theoretical discussion in Sect.~\ref{Methods} we use the Johnson filters as reconstructed by 
\citet{1990PASP..102.1181B} for backward compatibility because most of our published theoretical LCs have been based on this system.
Both the Swope and Johnson filters are used to study the sensitivity of the LC and color relations in Sect.~\ref{filter}.
All filters have been normalized to the Landolt system and the A0V star Vega \citep{landolt92}. 

As a reference model, and unless otherwise noted, the explosions originate from a progenitor with a zero-age main sequence mass of 
$M_{\rm MS}=5\unit{M_\odot}$ and solar metallicity, $Z_\odot$. 
At the time of the explosion, $\rho_c=2\times 10^9\unit{g}\unit{cm^{-3}}$ and  
$\rho_{\rm tr}=2.3 \times 10^7\unit{g}\unit{cm^{-3}}$ (Model 23). 
Two groups of models are used.
The first group is based on simulations with $\rho_{\rm tr}$ varied between $8-25\times 10^6\unit{g}\unit{cm^{-3}}$ \citep{HGFS99by02}. 
Models  08 -- 18 roughly constitute transitional and  subluminous objects (i.e., iPTF13ebh, SNe~1991bg- and 1986G-like, \citealt{hsiao2015,1993AJ....105..301L,1987PASP...99..592P}).
For the subluminous Models 08 and 16, additional models have been calculated with $\rho_c$ of $0.5$ and 
$1\times10^9\unit{g}\unit{cm^{-3}}$ (08r1, 08r2, 16r1, and 16r2).
Models 18 and 20 bracket the boundary between subluminous and normal-bright objects.   
For the normal-bright reference Model 23, the values of $M_{\rm MS}$ are chosen as $1.5$, $5.0$, and $7.0\unit{M_\odot}$ 
(23m2 -- 23m4) and $\rho_c$ is varied from $0.5-6 \times 10^9 \unit{g}\unit{cm^{-3}}$ (23d2 -- 5$^+$) \citep{2010ApJ...710..444H,tiara15}. 
In Fig.~\ref{LC_max}, we show variations{, among others,} in metallicity, $Z$.
However for the remainder of the analysis, we do not consider the effect of $Z$ because, { for DDT models,}
its influence on $B(t)$ and $V(t)$ can be expected { to be small, e.g. to be less than $0.02-0.03\unit{mag}$ at maximum light for $Z<Z_{\odot}$, based on synthetic spectra around maximum light \citep{hwt98,lentz01}. 
Note that the brightness in the $U$-band is very sensitive to $Z$
 \citep{hwt98,lentz01,2003ApJ...586.1199B,lentz01,Sauer08}. However, we cannot expect simple relations between $Z$ and $(U-V)$ because 
$U$ and the UV are equally sensitive to other quantities, such as the density structure at the photosphere, mixing, and, based 
on detailed spectral simulations by \citet{Gerardy}, to interactions with the environment and shells.
 These multiple, competing dependencies require a detailed analysis of observations based simultaneously on LCs, spectra, and  
line profiles \citep[e.g.,][]{tiara15}.}
Our models, their parameters, and observational properties are summarized in Table~\ref{table_mod}.

Representative abundances, velocity $v(m)$, and density structures $\rho(m)$ for a subluminous, SN~1991bg-like, supernova and the 
reference normal-bright SN~Ia are shown in Figs.~\ref{models} and \ref{models_rho}. 
The density and velocity structures depend only weakly on the free parameters of the model because both the initial WD structure and 
the energy released is determined by nuclear physics, particularly because the binding energy per nucleon is similar for both the 
elements in the iron group and the intermediate mass elements \citep{HGFS99by02,2006NewAR..50..470H}. 

The abundance structure is determined by the amount of burning during the deflagration phase, which regulates the pre-expansion of the 
WD and is parameterized by $\rho_{\rm tr}$. 
A smaller $\rho_{\rm tr}$ means more deflagration burning and a larger pre-expansion of the WD, leading to lower density burning during the 
detonation phase.
This results in an increased amount of intermediate mass elements, including \ce{Si} and \ce{S}, and explosive \ce{C}-burning products, 
including \ce{O}, \ce{Mg}, \ce{Ne}, rather than iron group elements, including \ce{Fe}, \ce{Co}, and \ce{Ni} 
\citep{HGFS99by02,2009MNRAS.399.1238H}. 
To first order, $\rho_{\rm tr}$ determines the \ce{{}^{56}Ni} production during the detonation phase 
(Figs. \ref{models} and ~\ref{LC_max}, { upper left plot}). 
The amount of nuclear burning during the deflagration phase causes some SNe~Ia diversity. 
Typically $0.2-0.35\unit{M_{\odot}}$ of high-density deflagration burning is required for the pre-expansion to agree with observations. 
Burning densities are sufficiently high to establish nuclear statistical equilibrium (NSE).
However, electron capture at densities above $\approx10^{9}\unit{g}\unit{cm^{-3}}$ results in a shift of the NSE from \ce{{}^{56}Ni} to 
mostly stable and long-lived isotopes of iron group elements \citep{brach00}. 
Thus, during the deflagration phase, the \ce{{}^{56}Ni} production also depends on the central density $\rho_c$ as shown in the 
{ right} panel of Fig.~\ref{models_rho}.
In normal-bright SNe~Ia, only some $10-20\%$ of the total \ce{{}^{56}Ni} is produced during the deflagration, whereas in subluminous 
SNe~Ia, deflagration burning dominates the total \ce{{}^{56}Ni} production.
This becomes evident in Figs. \ref{models} and ~\ref{LC_max} (upper left plot) and Table~2:  $\rho_c$ changes the
electron capture, and thus, the \ce{{}^{56}Ni} production during the deflagration. 
For example, the total $M(\ce{{}^{56}Ni})$ changes by a factor of 2 for Models 08 and 08r2 {  with 
 $\rho_c$ of $2$ and $0.5\times10^9\unit{g}\unit{cm^{-3}}$, respectively,}
 whereas for normal-bright SNe~Ia, the corresponding change is less than 15\%. { However, 
for both normal-bright and subluminous SNe~Ia, $\rho_c$ can have a profound impact on the LC 
and color relations, as discussed at the end of this section.}
 
After the explosion, the envelope enters a phase of fast expansion on time scales of seconds to minutes.
Adiabatic cooling results in a rapid drop of the temperature, $T$, in the envelope.
In regions without \ce{{}^{56}Ni}, the temperature drops well below $2{,}000\unit{K}$, resulting in low ionization and opacities \citep{h91}. 
Consequently, the photosphere in the optical and IR may not be formed at the outermost layers with the highest velocities. 
However, direct heating by $\gamma$-rays and energy diffusion quickly results in increasing opacities, producing a phase of 
{\it increasing} photospheric velocity \citep[see Fig.~1 in][]{HGFS99by02,2006NewAR..50..470H}. 
In our models, this phase lasts $\approx 1\unit{hour}$ for normal-bright SNe~Ia and $2-3\unit{days}$ for subluminous SNe~Ia.
Note that the duration of this phase is a very sensitive probe for the mixing-out of radioactive material \citep{HGFS99by02}.
Interaction with the surroundings or radioactive material close to the surface may nullify the effects of this phase.
Subsequently, on time scales of the \ce{{}^{56}Ni} decay and the diffusion time, the envelope becomes hotter during this {\it heating 
phase}.
A slight increase in the \ion{Si}{2} velocity within a few days of explosion was observed in the transitional object iPTF13ebh \citep{hsiao2015}.

The location of the photosphere, $R_{\rm phot}$, is dominated by geometrical dilution, namely the density of a mass element 
$\rho(m)\propto t^{-3}$, and the opacities drop with time.
The evolution of $R_{\rm phot}$ is shown in Fig.~\ref{models}. 
In SNe~Ia, the opacities are dominated by many overlapping emission lines and electron scattering, which form a quasi-continuum. 
The flux depends on the composition of the burning products, density gradients, velocity gradients, and temperature.
For a wide range of densities, velocity gradients, and chemical compositions, the transition from doubly- to singly-ionized iron group 
elements marks a regime of { rapid and steep} drops of the opacity with temperature, $T$.
This is due to lower ionization and excitation in addition to a shift of the emissivity towards longer wavelengths 
\citep[see Figs. 1 -- 4 in][]{h93}.
With a decreasing total \ce{{}^{56}Ni} production, the location of \ce{{}^{56}Ni} and the energy input becomes increasingly 
concentrated towards the inner, more opaque regions.
This results in a steeper temperature gradient and cooler envelope with smaller opacities. 
Thus, even for similar density structures, the photosphere recedes faster in mass coordinate $m$ with decreasing \ce{{}^{56}Ni} 
production. 
Observationally, the Doppler shifts of lines such as \ion{Si}{2}, \ion{S}{2} and \ion{Fe}{2} are expected to be decreasing with \ce{{}^{56}Ni}.
Moreover, the minimum photospheric velocities of elements such as Mg/O/Ne and the Si/S group and the line width of late-time
iron group elements will decrease with \ce{{}^{56}Ni}.

By about a month past the explosion, during the {\it semi-transparent phase}, the photosphere is formed well within the regions of iron 
group elements, with increasing energy input above the photosphere 
\citetext{see Fig.~10 in \citealp{h95} and Figs.~1 -- 3 in \citealp{h93}}.
This epoch corresponds to the change from absorption to emission features evident in the optical and most pronounced in the near- and 
mid-IR.
Both observations and models show the rise of emission features in the $H$  and $K$ bands produced by iron group elements \citep{1998ApJ...496..908W,HGFS99by02,Hsiao13}.   
For normal-bright and subluminous SNe~Ia the photosphere recedes to the \ce{{}^{56}Ni} layers around maximum light and 
$\approx10\unit{days}$ after maximum light, respectively.  
This change is also responsible for the rapid color evolution to the red $2-4\unit{weeks}$ past maximum light.    
After $30-40\unit{days}$, the photosphere of all models has receded to layers with $\rho \propto r^{-2...3}$ and NSE abundances. 
Thus, the time evolution of the density structure { and optical depth, $\propto t^{-3} $ and  $\propto t^{-2} $,} 
 cancels with the geometric dilution and consequently leads to similar conditions at the photosphere and similar colors. 
This is the physical basis for the Lira relation.
 Line transitions provide the coupling of the radiation field and the matter, and there are more line transitions in $B$ than in $V$. 
During the photospheric phase, the many transition lines of iron group elements acts as dark  ``absorption rods'' in front of a bright 
photosphere, because they transfer energy from the radiation field into the matter.  
Later on, these same lines act as bright ``emission rods''  showing up as emission features when most of the energy
 is deposited above the photosphere, because they transfer energy from the matter to the radiation field in the optically thin region.


\subsection{Light curves and their properties}\label{LC}

We want to discuss the $\Delta m_{15}$ relation, the color-stretch ($s_{BV}$) relation, the Lira relation, and CMAGIC in the context of our models.
$\Delta m_{15}$ is mostly sensitive to the diffusion time scale of the envelope, namely the optically thick layers.
The other relations measure the conditions in the photosphere, where the decoupling of photons occurs. 
The theoretical LCs properties near maximum light are shown in Fig.~\ref{LC_max}  and  LCs, $(B-V)(t)$ and $B(B-V)$ are shown in Fig.~\ref{LC_properties}.
For the theoretical comparisons, we use the time of the explosion as the $t=0$ reference.
For the comparison with observations, we switch to the more standard time of $T(B)_{\rm max}$ as the $t=0$ reference.

Historically, $\Delta m_{15}$ has been defined as the decline in the brightness between maximum and $15\unit{days}$ past maximum 
light \citep{p93}. One problem is the determination of the time of maximum light. 
Subsequently, this definition has been substituted by using template fitting of well-observed supernovae with different brightnesses
\citep{2006ApJ...647..501P} and stretching the time axis by a stretch factor $s$ \citep{1997ApJ...483..565P,1999ApJ...517..565P}.
This avoids the problem of diversity in the parameters potentially introduced in SNe~Ia samples with variable early and late-time 
coverage but good coverage around maximum light.  

For theoretical studies, the classical definition of $\Delta m_{15}$ is preferable because of its straight-forward interpretation as a 
measure of diffusion time scales and the expected variation caused by progenitors. 
 However, for subluminous SNe~Ia (Fig.~\ref{LC_properties}, upper left) the LC slopes flatten in the $11-14\unit{days}$ past 
maximum light. 
The faster evolution can cause the inflection point in $B$ and $V$ to occur before $15\unit{days}$ past 
maximum, causing a problem for measuring the pre-inflection smooth decline. 
Two sub-luminous SNe~Ia of different brightnesses can have the same $\Delta m_{15}$.
This effect was  pointed out by \citet{burns11} and is part of the motivation for developing the color-stretch parameter $s_{BV}$.      
\citet{2010ApJ...710..444H} dealt with this issue by first measuring the stretch in the time axis, $s$, of the light curve compared to a 
nominal $s=1$ light curve, in the same method as the stretch parameter from \citet{goldhetal01}. 
The modified decline rate $\Delta m_{15,s}$ is then measured as the magnitude decline from the time of peak to $s \times 15\unit{days}$ 
past maximum. We adopt the same method here.

{ In this paper, we consider $B$- and $V$-band LCs.
For  the detailed discussions of $\Delta m_{15,s}$, we focus on the $V$-band because it is frequently observed and can be used as well as a 
proxy for the bolometric LCs, $L_{\rm bol}$, around maximum light \citep{1992ARA&A..30..359B,2000A&A...359..876C}.
From theory, a smaller spread in
$\Delta m_{15} $ can be expected for a wide range of models and parameters because of its close link to $L_{\rm bol}$ \citep{HKWPSH96}.
Empirically,  \citet{p93} studied the $\Delta m_{15}$-relation in the $B$-, $V$-, and $I$-bands, and the spread in the $V$-band was found to be smaller than $B$ 
although the range of $\Delta m_{15}(B)$ has a wider range than $\Delta m_{15}(V)$. 
The wider range was important for low-accuracy LCs, but it is no longer 
as important with the advent of modern photometry.}

The $\Delta m_{15,s}$ parameters in $B$ and $V$ differ from the conventional $\Delta m_{15}$. 
However, for normal-bright and moderately subluminous  SNe~Ia $\Delta m_{15} \approx \Delta m_{15,s}$, which enables a link to 
previous literature. 


\subsubsection{Light curve properties near maximum light}

\begin{figure}   
\includegraphics[angle=360,width=0.98\textwidth]{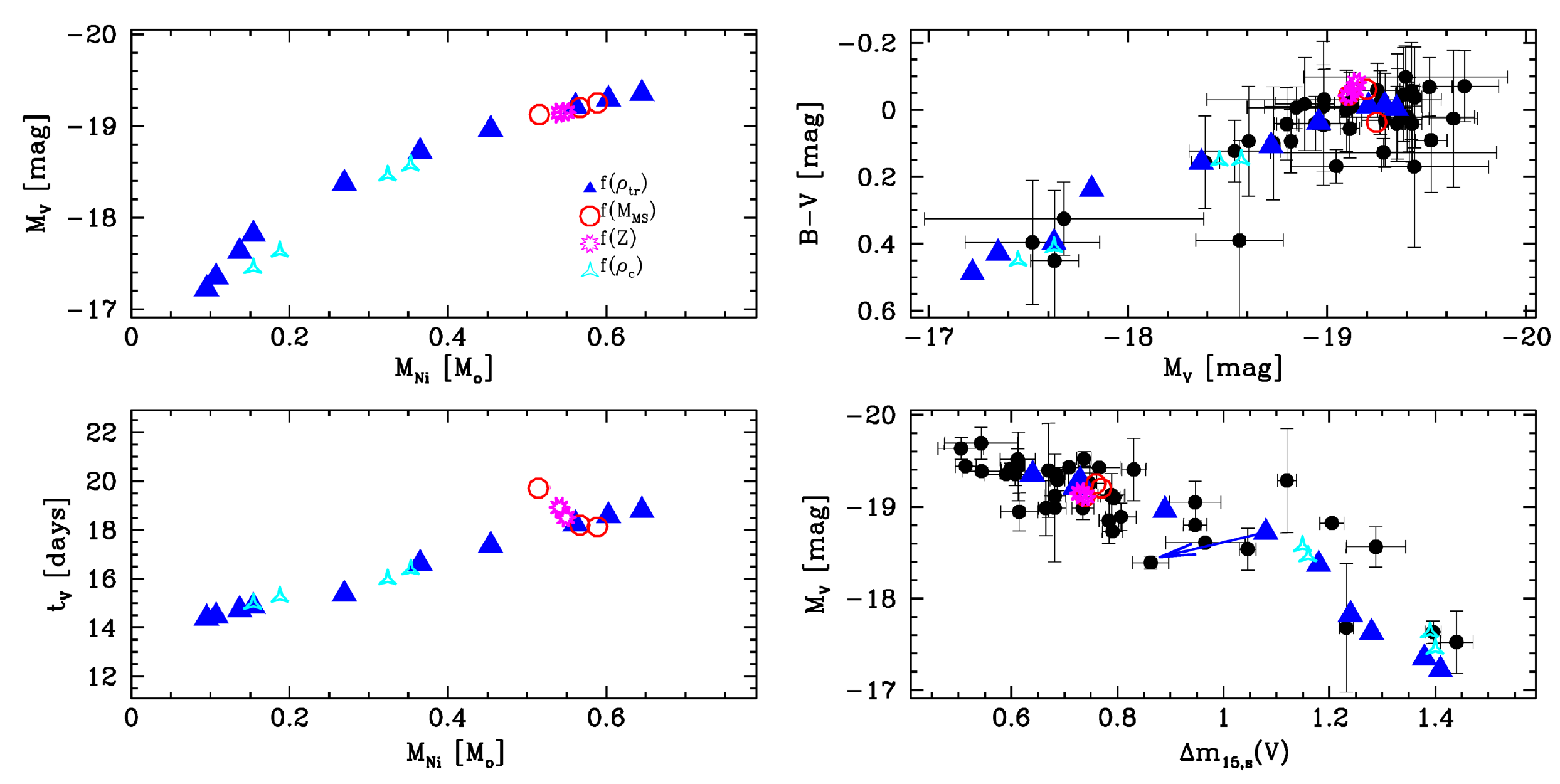}
\caption{
Light curve properties of our reference model with variations in the basic parameters $\rho_{\rm tr}$ (blue triangles),  
 $M_{\rm MS}$ and metallicities $Z$ (red open circles and stars). 
In addition, the loci are given for $\rho_c$ for models 08 and 16 (cyan triangles).
We show the maximum brightness $M_V$ { versus \ce{{}^{56}Ni} mass, $M_{\rm Ni}$} (upper left), the rise times $t_V$  {versus $M_{\rm Ni}$} (lower left), the $(B-V)$ color at maximum  versus peak $M_{V}$} (upper right), and  {peak $M_{V}$ versus $\Delta m_{15,s}(V)$ (lower right). 
A major uncertainty in the theoretical value of $\Delta m_{15,s}$, at the $\approx 0.1\unit{mag}$ level, is caused by a slow evolution in 
brightness around maximum light. 
We give models with $\rho_{\rm tr}=0.8-2.7\times10^7\unit{g}\unit{cm^{-3}}$ (blue triangles), $M_{\rm MS}=1.5,$ $3.$, and 
$7.\unit{M_{\odot}}$ (red circles), and $Z=0.1$ and $10^{-2}\unit{Z_{\odot}}$ (magenta star). 
The blue arrow gives the shift of Model 18 if we mix the inner $8{,}000\unit{km}\unit{s^{-1}}$. 
	In addition, we show the loci of 43 CSP~I supernovae with pre-maximum data using redshift distances and empirical reddening corrections (see text).}
\label{LC_max}
\end{figure}

Some of the well-established LC properties are shown in Fig.~\ref{LC_max}.
To first order, the LCs of SNe~Ia are powered by the radioactive decay of \ce{{}^{56}Ni} \citep{ColgateMckee69}. 
More \ce{{}^{56}Ni} increases the brightness of the SN, which is the basis of Arnett's rule \citep{1979ApJ...230L..37A}. 
Compared to the instantaneous energy deposition by radioactive decay ($\alpha = 1$), the actual SN brightness is expected to be larger by a factor of 
${ {\alpha}}\approx 1.2-1.3$  with a dispersion ${ \sigma {\alpha}} \approx 0.2$ \citep{hk96,HGFS99by02}. 
The basic dependencies can be understood in terms of the rapidly dropping { Rosseland and flux-mean} opacities at low temperatures.
An increased temperature produces higher opacities and, thus, longer diffusion time scales. {Note that the opacities 
are dominated by overlapping lines forming a `quasi-continuum', but Thomson scattering is very important in closing the `gaps' in frequency
space (see \citealt{karpetal77} and Fig. 3 in \citealt{h93}).}
As a consequence, the rise time to maximum light increases, and after maximum the envelope takes longer to release the extra stored 
energy, causing a smaller $\Delta m_{15,s}$ and $\Delta m_{15}$ \citep{HKWPSH96,nughydro97,umeda99,2000ApJ...530..744P,kasen09}. 
A necessary condition is that the expansion rate of the envelope is sufficiently low so that adiabatic cooling does not over-compensate for 
the energy production by \ce{{}^{56}Ni}.  
The $\mch$ WD models and the observations both show a rapid drop in maximum brightness at $\Delta m_{15}(B)\approx1.5$ and 
$\Delta m_{15}(V)\approx1.2$, which we call the ``$\Delta m_{15}$-cliff'' (see Fig.~\ref{LC_max}, lower right). 
From theory, this drop is caused by a sufficiently low photospheric temperature, allowing the SN to enter a regime of quickly-dropping 
opacity with temperature as discussed in Sect.~\ref{bgm}.

The tightness of the brightness decline relation requires similar \ce{{}^{56}Ni} distributions.
From physics, if { different} SNe have a variable amount of mixing, the tight $\Delta m_{15}$ relation will be destroyed
\citep{HKWPSH96,kasen09}, and thus, mixing needs to be avoided within DDT models, as shown with the shift in Model 18 when mixing 
is allowed (Fig.~\ref{LC_max}, lower right). 

We note that the existence of a  $\Delta m_{15}$ relation holds up for virtually all { thermonuclear} explosion scenarios 
as long as there is an excess of stored energy. 
However, the position of the ``$\Delta m_{15}$-cliff'' must be expected to change with the total WD mass, $M_{\rm WD}$, because for 
similar opacities the diffusion time scales as the square of the total mass, ${M_{\rm WD}}^2$.
Alternatively, it has been suggested that SNe~Ia ``beyond the cliff'' can be attributed to a different explosion scenario than normal-bright 
SNe Ia \citep{2007A&A...476.1133F,kromer2010}. {Very recently, Blondin et al. (2017) presented a paper based on both classical 
and pulsating delayed-detonation (PDD) and sub-$M_{Ch}$ models. They found that their $M_{Ch}$ mass models follow the Phillips relation
only for values of $ \Delta m_{15}(B) < 1.38~mag$ and $ < 1.25~mag $ for DDT and PDD models, respectively, which is  well before the ``$\Delta m_{15}$-cliff''. Subsequently,  
their theoretical relations turn sharply and almost orthorgonal to the observed Phillips relation. A similar anti-correlation 
was found by \citet{PE00} and later vanished 
\citep{2000ApJ...530..744P,PE01}.  Furthermore, sub-$M_{Ch}$ models of \citet{Blondin17} 
show $\Delta m_{15}(B) $ values up to $\approx 1.6 mag$ and with a steep rate of decline before showing a similar turn than their $M_{Ch}$ 
models. From this finding, the authors argue that $\Delta m_{15}$ can only be reproduced with sub-$M_{Ch}$ models at faint end of 
SNe~Ia. In contrast, and based on spectro-polarimetry of the subluminous SN~2005ke, \citet{patat12} concluded
that subluminous SNe~Ia may either be understood within  $M_{Ch}$ or merger scenarios of two WDs.} 
Here we will show, that different scenarios will also produce color evolutions for similar brightness inconsistent with the observations. 

In cosmology, the $\Delta m_{15}$ relation for normal-bright SNe~Ia plays a key role for standardizing SN~Ia brightnesses for cosmological studies.
To first order the brightness increases with the production of \ce{{}^{56}Ni}. 
However, variations in the progenitor system such as $M_{\rm MS}$ and metallicity of the progenitor will unavoidably introduce 
a color dispersion, diminishing our ability to distinguish between intrinsic color changes at approximately the $0.1\unit{mag}$ level and 
reddening by the interstellar medium (ISM).
This effect is shown in the upper panels of Fig.~\ref{LC_max}. 
In principle, variations in the rise time to maximum light of SNe~Ia may be employed to disentangle intrinsic color changes and reddening 
by the ISM (Fig.~\ref{LC_max}, lower left), however this procedure relies on the homogeneity of SNe~Ia explosion scenarios.

Secondary parameters such as $\rho_c$ and $M_{\rm MS}$ also introduce diversity in the $\Delta m_{15,s}$ relation, and the theoretical 
dispersion is approximately $0.05\unit{mag}$ for normal-bright SNe~Ia (Fig.~\ref{LC_max}, lower right).
Varying the $\rho_c$ changes the amount of \ce{{}^{56}Ni} produced during the deflagration phase, which becomes increasingly 
important for dimmer SNe~Ia. 
In comparison, for our subluminous Model 08, the photosphere recedes much faster in velocity space than for a normal-bright SN~Ia 
and, consequently, energy can diffuse into the photosphere before maximum light. 
Considering changes in the $\rho_c$ for subluminous SNe~Ia, Models 08r1 and 08r2 are brighter by about $0.5$ and $0.25\unit{mag}$ 
than Model 08, respectively, but still show a similar $\Delta m_{15,s}$ (Fig.~\ref{LC_max}, lower right).\footnote{The corresponding 
model with $\rho_c =4 \times 10^9\unit{g}\unit{cm^{-3}}$ (not included in Table~\ref{table_mod}) is dimmer by about $0.3\unit{mag}$.}
A spread of similar size has been observed \citep{contreras2010,2011AJ....142..156S,Stritzinger14}.

Variations in $M_{\rm MS}$  and $Z$ of the progenitor influence the expansion rate of the ejecta and, thus, the heating at the 
photosphere \citep{hwt98}. 
Moreover $Z$, or more precisely the \ce{{}^{22}Ne} abundance, reduces the \ce{{}^{56}Ni} production, for $Z \leq Z_\odot$ in 
particular \citep{hwt98,timmes03}.
The main consequence of varying these parameters is a spread in the intrinsic color at maximum light, $(B-V)_0$, of about 
$0.1\unit{mag}$ (Fig.~\ref{LC_max}, top right). 
If the interstellar reddening $E(B-V)$ correction is determined under the assumption that $(B-V)_0$ is identical for SNe~Ia with similar 
$\Delta m_{15,s}$, this may lead to significant distance errors. 
Unfortunately, the effect of these secondary parameters on the \ce{{}^{56}Ni} production and of reddening 
on the brightness are along the same offset  vector in a plot of the $\Delta m_{15,s}$ relation, making them hard to distinguish between.
If the dust properties vary with redshift, this effect may introduce a major bias into high-precision cosmology using SNe~Ia.

We note that for $\mch$ models with the same $\rho_{\rm tr}$ and $M_{\rm MS}$ but different $\rho_c$, spectra and LCs will form in 
almost identical density, velocity, and chemical abundance layers up to about $1-2\unit{weeks}$ after maximum light, since the main 
differences show up in the central regions only. 
However, even for those ``twins'', the maximum brightnesses $M_B$ and $M_V$ may differ by about $0.05\unit{mag}$ in normal-bright 
SNe~Ia and more than $0.7\unit{mag}$ in subluminous SNe~Ia. Therefore, we stress the importance of observing SNe~Ia up to $
+40\unit{days}$ after maximum light to probe the influence of $\rho_c$ and to obtain accurate absolute magnitudes (see, e.g., Gall et al. 2017, submitted).

For high-precision cosmology, the spread in the observed $(B-V)$ color must be expected due to variations in the progenitor system, 
explosion scenario, and properties of the interstellar medium. 
This then demands the additional use of the time-dependence of color information.
Reddening by the ISM can be assumed to be time-independent, whereas variations in the progenitor system or explosion scenario can 
be expected to produce time-dependence in color.  
{ Some of the intrinsic variations may be detected by differential light curve analyses, for example by using the peak-to-tail ratio, namely 
a relative offset in the monochromatic LC-tail of two SNe~Ia with similar $\Delta m_{15,s}$ \citep{2010ApJ...710..444H}. 
Such an off-set can be produced by differences in $\rho_c$. Around maximum light, central $^{56}$Ni does not contribute to the LC due to the long diffusion 
time scales, but it does add flux during the tail of the LC (see Fig.~\ref{p2t}). Such variations have been observed and analyzed, and being monochromatic, should 
not depend on { reddening by} the  ISM  \citep{sadler12,h13}.}    
However, the peak-to-tail ratio neglects color information, which we will address in the next section.


\subsubsection{The $(B-V)$ color evolution and Lira relation}\label{lira}

\begin{figure}   
\includegraphics[angle=360,width=0.98\textwidth]{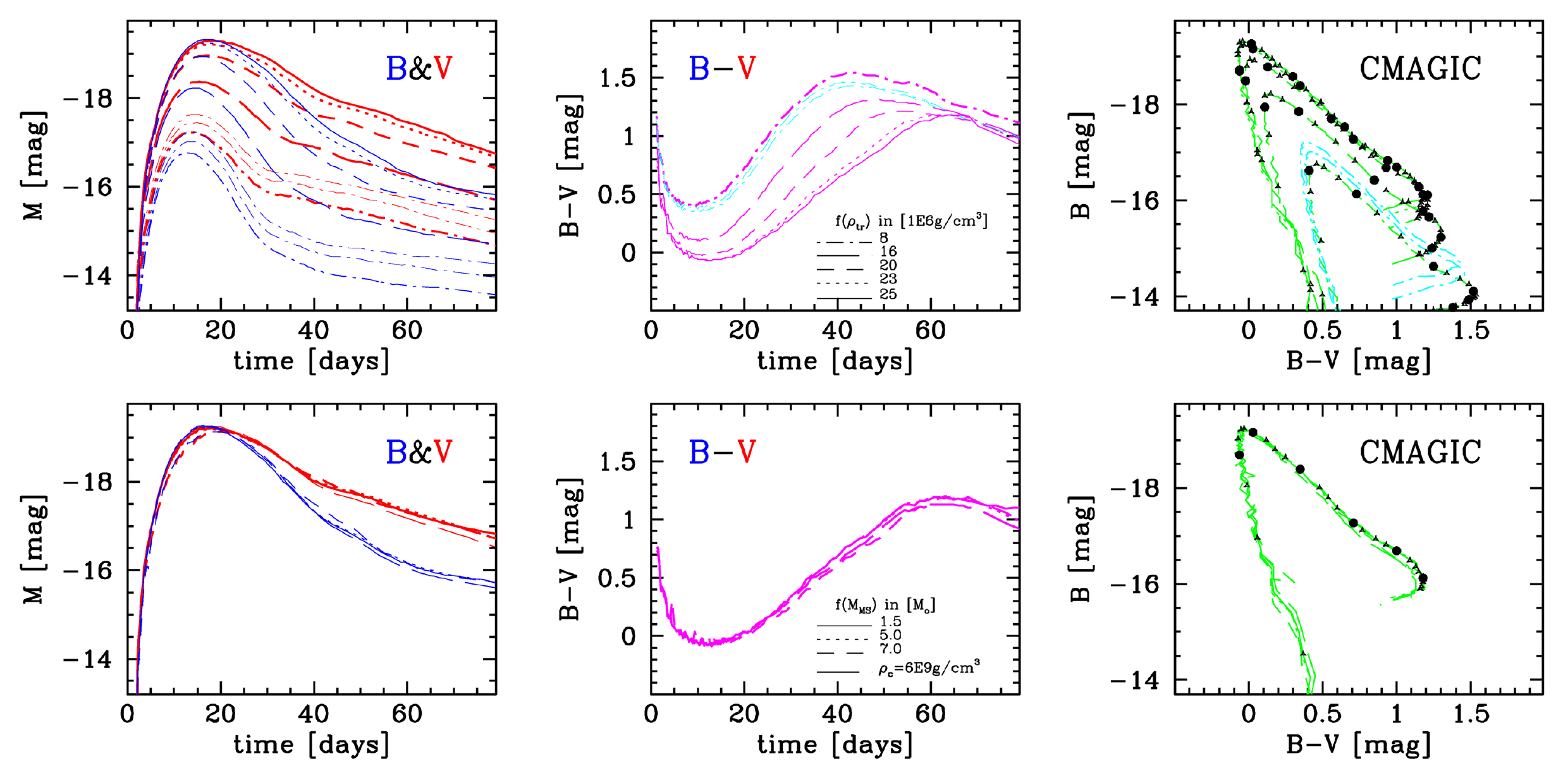}
\caption{
Theoretical $B$- and $V$-band light curves, color evolution $(B-V)(t)$, and the color-brightness diagrams, CMAGIC \citep{cmagic03}. 
{ The time $t$ is given in days past the explosion.} 
In the upper panels, $\rho _{\rm tr,6}$ (in $10^6\unit{g}\unit{cm^{-3}}$) has been varied to produce normal-bright SNe~Ia (Models 23 and 
25), a model at the edge of the $\Delta m{15}$-cliff (Model 20), and subluminous SN~1986g-like (Model 16) and SN~1999by-like (Model 
08) models \citep{2006NewAR..50..470H}. 
In addition, SN~1991bg-like models are given with central density varied with $\rho_c=0.5$ and $1.1\times10^9\unit{g}\unit{cm^{-3}}$ 
(thin, dot-slash).  
In the lower panels, we varied secondary parameters for our reference model, originating from stars with 
$M_{\rm MS}$ of $1.5$, $5.$, and $7.\unit{M_{\odot}}$ and with $\rho_c=2.$ and $6.\times10^9\unit{g}\unit{cm^{-3}}$. 
In the CMAGIC plots (right), the big circles (dots) and small triangles (black) mark the times with $\Delta t$ of $10$ and $2.5\unit{days}$, 
respectively. 
In the $B(B-V)$ curves, models evolve along the lines starting at the lower left. { Throughout this paper we refer to the path towards the first rapid change (1st leg), the path towards the second rapid change (2nd leg), and past the last point of rapid change (3rd leg).}
The evolution is characterized by a fast change during the first $10\unit{days}$ and a very slow evolution after about $40-50\unit{days}$. 
Note that for subluminous SNe~Ia (upper left plot), the LC slopes flatten some $11-14\unit{days}$ past maximum light. 
Two models with different LCs may have the same $\Delta m_{15}$. 
To avoid this problem, we use the stretch-corrected $\Delta m_{15,s}$ with the LC stretch-parameter $s$ \citep{1997ApJ...483..565P} 
and calculate the decline rate over a duration $t_s = s\times15\unit{days}$. 
We use the label {\sl time} when referring to the time since the explosion.
For a detailed discussion, see text.
}
\label{LC_properties}
\end{figure}

In contrast to the $\Delta m_{15,s}$ relation, which is most sensitive to variations in the optically thick regions of the envelope, colors are determined mostly by the conditions at the photosphere. 
The empirical relations $s_{BV}$ and the Lira relation both rely on color information and color evolution to 
characterize a SN~Ia. 
CMAGIC is a hybrid of these relations that directly links color information to brightness. 
To understand these relations, their stability, and possible systematics, we need to analyze the physics which governs the color evolution 
$(B-V)(t)$ (see Fig.~\ref{LC_properties}).
We have broken down the evolution into four phases that are dominated by slightly different physics.

{\sl Phase 1:} 
The early color evolution is characterized by a decrease in $(B-V)$, which ends $11-14\unit{days}$ after the explosion and well before 
maximum light. 
This blueing is caused by radioactive decay in the inner \ce{{}^{56}Ni} region below the photosphere and diffusion by low energy photons. 
Due to geometrical dilution, the photosphere recedes toward the heat source and $(B-V)$ becomes bluer. 
Heating by \ce{{}^{56}Ni} is intense compared to the later \ce{{}^{56}Co}-dominated regime because the specific energy deposition per 
time is higher by a factor of $\approx 5$. 
The short \ce{{}^{56}Ni} lifetime, $8.80\unit{days}$, governs the evolution time scales for all models.  

{\sl Phases 2 and 3:} 
After the initial heating phase, the color evolution diverges between normal-bright and subluminous SNe~Ia, with the boundary between 
the two groups separated by the ``$\Delta m_{15}$-cliff''.

{\sl Phase 2 for normal-bright SNe~Ia:} 
Within the brightness range for normal-bright SNe~Ia \citep[see Figs.~10 -- 11 in][]{h95}, $(B-V)$ color is similar because the energy 
distribution in the optical, including both $B$- and $V$-bands, is dominated by ionization and excitation effects.
In the days before maximum light, the photospheric temperatures are similar to an A0V star such as Vega with  $(B-V) \approx 0$.
Second-order effects are caused by line-blocking: the recombination of doubly-ionized iron group elements causes selectively higher 
line-blocking in the $B$ band compared to in the $V$ band. 
Approaching maximum light, the colors become redder despite the increasing luminosity. 
The increase in overall heating is compensated by the increase of the photospheric radius. 
Because the photosphere is formed in layers with steep density gradients, the energy density at the photosphere decreases and $(B-V)$ 
becomes redder prior to maximum light. 
The competition of these two effects leads to a wide minimum in the color evolution plot. 

{\sl Phase 3 for normal-bright SNe~Ia:}
After maximum light, the color continues to redden because of the decreasing luminosity. 
The color evolution is similar among normal-bright SNe~Ia because the quasi-continuum forms in regions with very similar chemical 
abundances, namely in the inner \ce{{}^{56}Ni} region, and with flat density gradients, which compensate for the geometrical dilution with 
time.  

{\sl Phase 2 for subluminous SNe~Ia:}
The energy balance is dominated by the heating in layers of singly-ionized iron group elements and intermediate mass elements well 
before maximum light. 
As a result, the temperature at the photosphere depends on the total production of \ce{{}^{56}Ni} even a few days before maximum. 
Less heating by \ce{{}^{56}Ni} decay means a cooler photosphere and an earlier-occurring and redder minimum in the $(B-V)$ color 
evolution plot.

{\sl Phase 3 for subluminous SNe~Ia:} 
Subsequently, $(B-V)$ increases and the envelope becomes redder because of the declining energy input. 
Even after maximum light, the photosphere is formed in layers of intermediate mass elements for a period of time that depends on the 
total amount of \ce{{}^{56}Ni} produced in the SN, causing diversity in the $(B-V)(t)$ plot. 
Note that the photosphere recedes {\it faster} with a smaller \ce{{}^{56}Ni} production, but the corresponding effect of an increased 
production of intermediate mass elements is dominant (see Fig.~\ref{models}).
By about $30\unit{days}$ post-explosion, the photosphere has entered the NSE region even in very subluminous, SN~1991bg-like, 
objects, as can be seen in Fig.~\ref{models} \citep[also see Figs.~13 -- 14 in][]{h95}.

{\sl Phase 4:}
For the entire range of models, the physics after maximum light is rather similar because the photosphere forms in layers dominated by 
iron group elements and, with time, a larger fraction of the energy is deposited in the optically thin region. 
Lines from atomic transitions act as bright ``emission rods'' in front of a decreasing flux from the photosphere.
The larger line-blanketing { and more lines} in the $B$-band compared to $V$-band means the $(B-V)$ color becomes bluer with time.
{ Note that during the transition from the photospheric to { nebular} phase, very strong lines are still optically thick.
Most spectra and spectral features in the optical and NIR are formed by multiple overlapping lines on top of a  
quasi-continuum during the transitional up to the nebula phase, i.e., one month to a year after maximum light
\citep{h95,1996MNRAS.278..763B,gerardy07,2008PASP..120..135B,2008MNRAS.386.1897M,tiara15,2015ApJ...798...93T}. 
The dominance of the continuum for the broad-band flux also becomes evident from observations ranging from $3\unit{weeks}$ to $3\unit{months}$ after maximum (see Figs. 17 to 26 shown in      
\citealt{2008MNRAS.386.1897M}). Empirically, the nature of the late-time optical features as a result of many overlapping lines
becomes also apparent in low-velocity thermonuclear explosions such as SNe~Iax. Due to the low Doppler shift, 
the broad features known from SNe~Ia frequently appear as multiple-components \citep{Stritzinger14,2015A&A...573A...2S}.}

Comparing subluminous to normal-bright SNe~Ia, the opacities and ionization levels are lower, which results in the photosphere receding 
faster.
Subluminous SNe~Ia enter {\sl Phase 4} earlier than normal-bright SNe~Ia. 
The transition between {\sl Phase 3} and {\sl Phase 4} is correlated with the post-maximum inflection points in the $B$- and $V$-band 
LCs.\footnote{We note that the timing of the secondary IR maximum does not directly correlate with the inflection point of the LC.  
{ The primary IR maximum is a result of the evolution of the bolometric
luminosity around maximum light similar to $B$ and $V$.}}
The secondary maximum in the IR colors mark the time of maximum radius of the photosphere $R_{\rm phot}$. 
In the IR, the source function is in the Wien's limit, i.e. the flux $\approx T{R_{\rm phot}}^2$. 
The IR flux increases to a secondary maximum when $R_{\rm phot}$ reaches is maximum value. 
The photospheric radius decreases when the density gradients become smaller than $\propto r^{-3}$ or it cannot compensate for the 
opacity decreasing as the temperature drops \citep{hkw95}. 
With decreasing total \ce{{}^{56}Ni} production, the time of the secondary IR maximum shifts towards the primary maximum and 
eventually merges \citep{hkw95}.
{ We add that $B$- and $V$-band LCs do not show secondary maxima because the energy distribution peaks longward of $V$ and the 
emissivity depends exponentially on $T$.}
During the time period where the Lira relation holds, $\approx 30-60\unit{days}$ after maximum light ({\sl Phase 4}), the spectra of the entire 
range of models form in the regions with iron group elements, which have similar density structures and abundances. 
To first order, all models show a rather similar color evolution (see Fig.~\ref{LC_properties}). 

\begin{figure}[ht]
$
\begin{array}{c}
\includegraphics[width=0.96\textwidth]{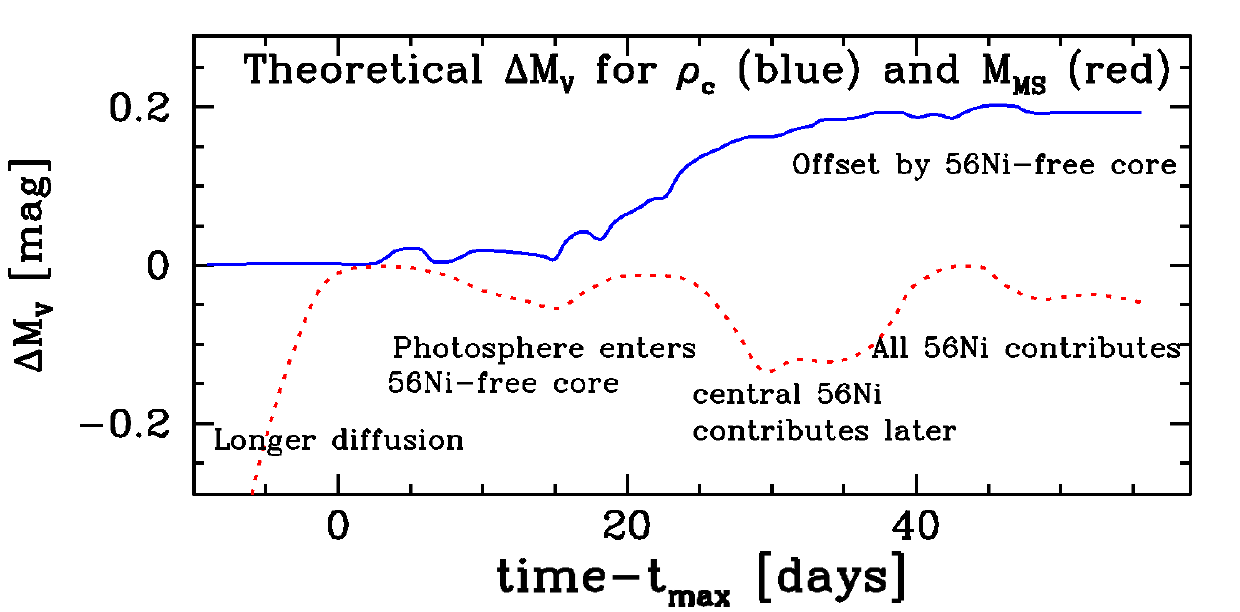}
\end{array}$
\vskip -5pt
\caption{
Variations in the light curves due to changes in the progenitor as a function of time past maximum $t_{max}$, namely its $M_{\rm MS}$ and its central density $\rho_c$, are hardly 
visible in overall light curves (lower plot of Fig.~\ref{LC_properties}, lower left plot) but become apparent when plotting the differences.
On the left models with $\rho_c=6\times10^9\unit{g}\unit{cm^{-3}}$ and  
$M_{\rm MS}=7\unit{M_{\odot}}$, respectively, relative to Model 23 with $\rho_c=2\times10^9\unit{g}\unit{cm^{-3}}$ and 
$M_{\rm MS}=5\unit{M_{\odot}}$.
}
\label{p2t}
\end{figure}

Models with various $M_{\rm MS}$ and $\rho_c$ are shown in the lower panels of Fig.~\ref{LC_properties}.
As discussed in Sect.~\ref{LC}, $\rho_c$ significantly affects the \ce{{}^{56}Ni} production during the deflagration phase.
For normal-bright SNe~Ia models, this produces the peak-to-tail relation 
\citep[see Fig.~\ref{p2t} and][]{2010ApJ...710..444H,sadler11,h13}.
The diffusion time scales are comparable to or longer than the rise time to maximum. 
As a result, \ce{{}^{56}Ni} decay in the central layers hardly contributes to the luminosity up to $7-10\unit{days}$ past maximum light. 
{ Thus}, two supernovae of different $\rho_c$ may have similar early LCs, but the LC ``tails'' are offset (Fig.~\ref{p2t}).
For our models of a subluminous SNe~Ia (Model 08), the more centrally located \ce{{}^{56}Ni} will result in additional heating of the 
photosphere as early as maximum light (Figs.~\ref{models_rho} and \ref{LC_properties}) and a lower $\rho_c$ also provides additional 
heating in the subluminous models. 
Although the color evolution of Model 08 is similar to that of Model 08r2 (with a lower $\rho_c$) during {\sl Phases 1} and {\sl 2}, the 
subsequent color evolution of Model 08r2 resembles that of a sub-luminous, SN~1986G-like object with a higher $\rho_{\rm tr}$ and 
corresponding to our Model 16.

By about $40\unit{days}$, the diffusion time scales are less than $1-2\unit{days}$ for our models.
The $(B-V)$ color evolution converges to the Lira relation at about $40\unit{days}$ after maximum light. 
Compared to our reference, models with lower $\rho_c$ are bluer by $\approx 0.2\unit{mag}$ for SN~1991bg-like SNe~Ia and 
$\approx 0.1\unit{mag}$ for normal-bright SNe~Ia.
The underlying physics produces the observed uniform color evolution in normal-bright SNe Ia and a higher dispersion in the color 
evolution of subluminous SNe Ia.

The evolution in magnitude-color space, $B(B-V)$ or CMAGIC, is shown together with time stamps indicated by black dots in 
Fig.~\ref{LC_properties}. 
The slow brightness evolution in $B$ and $V$ around maximum light and after $50\unit{days}$ produces characteristic ``knees'' in 
the CMAGIC curve. 
The CMAGIC method compresses the time evolution where the conditions at the photosphere change as the SN goes through the 
different physical regimes corresponding to {\sl Phases 1 -- 3}.

{\sl Phase 1}, approximately up to two weeks past explosion, is characterized by a rapid heating. 
During this time, the photosphere is formed in layers of intermediate mass elements and very similar ionization levels for all models. 
For all normal-bright models, the evolution of $B(B-V)$ is smooth after the first ``knee'' because the density
{ profile is shallow} in
\ce{{}^{56}Ni} region. 
For the subluminous SNe~Ia (Models 08 and 16), the photosphere is mostly formed in the region of unburned and intermediate mass 
elements, showing a red ``bulge'' at $(B-V) \approx 1.2$ mag and $1.3$ mag when entering the \ce{{}^{56}Ni} layers. 
The location of the two ``knees'' is shifted along the lines $l_{1,2}$:
\[
l(B-V)_{1,2} \approx C_{1,2} +D_{1,2} (B-V),
\]
where $C_1=-19.3$, $D_1=6.1$, $C_2=-16.1$, and $D_2=5.7$.
We note that the shift of $l_1$ is sensitive to the maximum brightness of the SN. 
The location for the second ``knee'' depends on $\rho_c$ because the density influences the amount of central \ce{{}^{56}Ni} and, thus, 
the heating of the late photosphere.  

\begin{figure}   
\includegraphics[angle=360,width=0.98\textwidth]{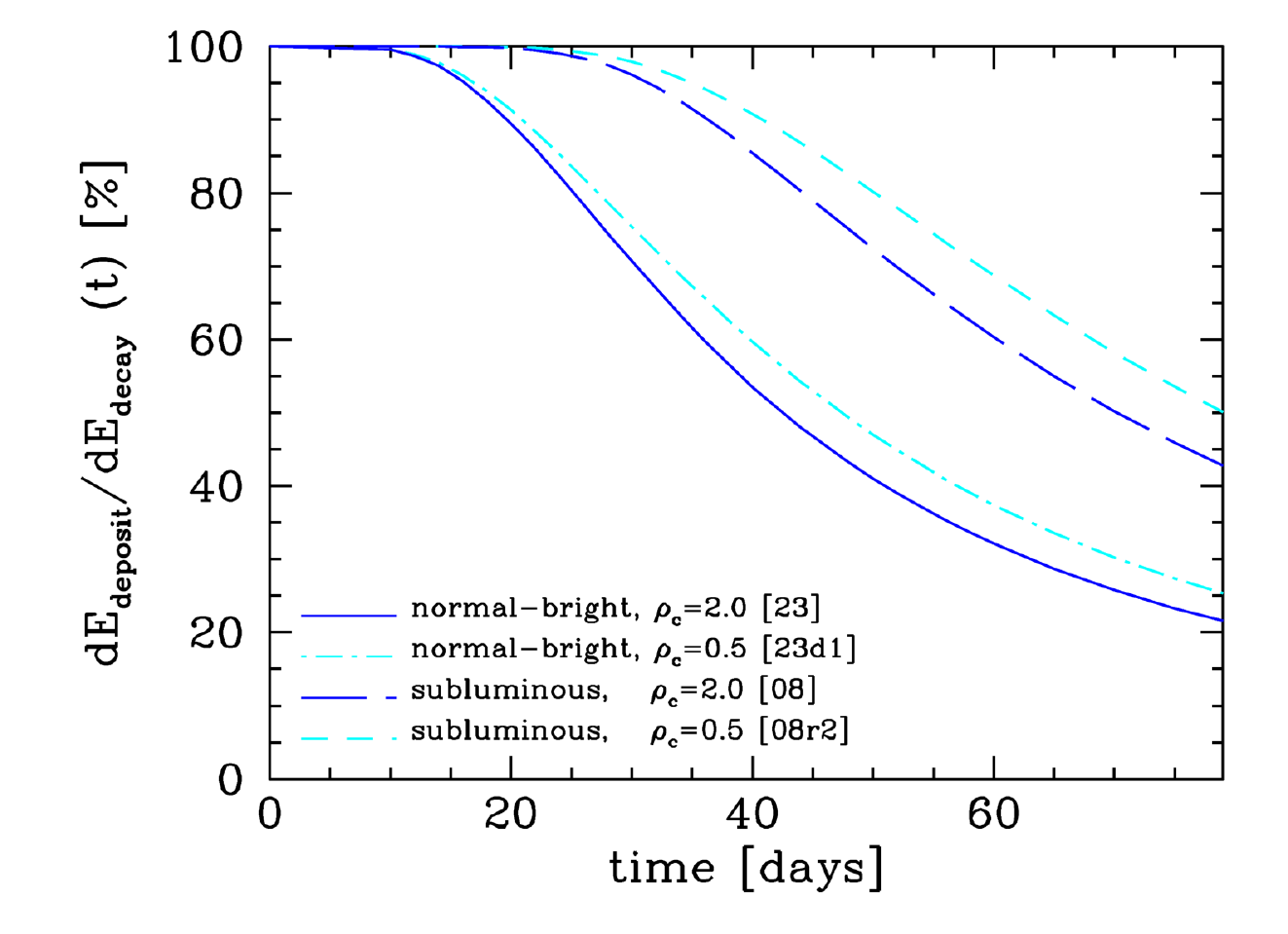}
\caption{{ Relative energy deposition $dE_{\rm deposit}/dE_{\rm decay}$ by radioactive decay of $\ce{{}^{56}Ni}\rightarrow\ce{{}^{56}Co}\rightarrow\ce{{}^{56}Fe}$ via
electron capture and $\beta^+$ emission for a normal-bright and subluminous model with $\rho_c$ of $0.5$  and $2\times10^9\unit{g}\unit{cm^{-3}}$.  
In brackets, we give the model names corresponding to Table 1. 
 For both normal-bright and subluminous models, lower $\rho_c$ results in more \ce{{}^{56}Ni} (Table 1) and \ce{{}^{56}Ni} located closer to the center 
(Fig. \ref{models_rho}) and, thus, larger $dE_{\rm deposit}/dE_{\rm decay}$. More heating and the short diffusion compared to the expansion time scales
in the weeks after maximum light means both bluer colors and larger brightnesses (see text and Fig. \ref{LC_properties} ).}
}
\label{gamma}
\end{figure}

{ The origin of the  peak-to-tail ratio and the dispersions during the Lira relation phase in $(B-V)(t)$ and on the off-set in
the brightness-color relations can be attributed to $\rho_c$ as the most dominant contributor.  
As discussed above, lower $\rho_c$ means more central \ce{{}^{56}Ni}, which mostly contributes to the LCs after maximum because the photon diffusion time scales are larger than the rise time to maximum. Past maximum, more central \ce{{}^{56}Ni} results 
in a larger relative energy deposition rate $dE_{\rm deposit}/dE_{\rm decay}$ by radioactive decay.   
An example of this effect can be seen for the normal-bright Models 23d1 and 23 and the most subluminous Models 16r2 and 16, with $\rho_c=0.5$ and $2\unit{g}\unit{cm^{-3}}$,
respectively.
In Fig. \ref{gamma}, $dE_{\rm deposit}/dE_{\rm decay}$ is shown for a normal-bright and a subluminous model. 
Up to about maximum light, most of the decay energy is trapped but $\gamma$-ray photons in the outer layers
can escape increasingly with time.
$\rho_c$ changes the energy deposition by $<1\% $ about maximum light but $\approx 20\%$ after some 30 to 40 days after the explosion.
The consequence is a shift in peak-to-tail brightness by $\approx0.2\unit{mag}$ (Fig.~\ref{p2t}) and
$\approx 0.3\unit{mag}$ over the entire $\rho_c$ range discussed (see Table 2, \citealt{sadler12}, and \citealt{h13}), and more heating leads to 
bluer colors during the Lyra relation. For our subluminous model, the difference in color is $\approx0.2\unit{mag}$.
 For normal-bright SNe~Ia, the color effect is smaller, $\approx0.1\unit{mag}$, because most of the \ce{{}^{56}Ni}
is produced during the detonation phase.

{ Note that a change in $Z < Z_\odot$} hardly affects the \ce{{}^{56}Ni} production    
or the explosion energy.  $M_{\rm MS}$ changes the geometrical dilution, however, some weeks after maximum, most of the envelope with \ce{{}^{56}Ni}
is optically thin for $\gamma$-rays and the off-set in brightness becomes small (Fig.~\ref{p2t}).}


\subsection{Limitations of the models}\label{limits}

Model predictions are typically stable if they depend on global quantities, such as the relationship between luminosity and \ce{{}^{56}Ni} 
production, if they depend on self-similar solutions, such as the structure of a self-gravitating gas ball supported by degenerate 
electron gas, or when using differential changes of given quantities, such as differences between LCs. 
However, model predictions have their own set of uncertainties. 
The determination of $t_{\rm max}(B-V)$ and $t_{\rm max}(V)$ is particularly sensitive to uncertainties in the explosion model. 
Examples include macroscopic mixing and microphysics, namely incomplete or inaccurate atomic cross sections, missing atomic line 
transitions, and the use of ``supermodels'' employed in all modern non-LTE simulations 
\citep{h95,lentz01,dessart14,2015MNRAS.454.2549B,2015ApJ...798...93T}.
Errors due to microphysics build up with time. 
Early on, the radiation field is thermalized in most parts of the envelope and the mean intensity, $J$, is close to the local Planck function, 
$B(T,\lambda)$, which is independent of the atomic physics.  
At maximum light,{ $J$ differs from $B$}  by a factor of about $3-5$ even at deep layers and, increasingly, the energy input by radioactive decay is 
above the photosphere \citep[Fig. 3 in][]{h95}. 
The lack of included forbidden transitions and missing atomic data affects both the spectra and colors{, in particular, beyond 60 days after the
explosion \citep{h95,dessart14,2015MNRAS.454.2549B,2015ApJ...798...93T}. At late times, the lack of cooling channels results in   
hotter layers and too fast of a slope in  $(B-V)(t)$, as we will see below.}
Discrepancies at early times are most likely attributed to shortcomings of our explosion models and may hint at some limited amount of 
mixing during the deflagration phase or interactions with material from the progenitor system and the nearby environment. 


\subsection{Influence of system response functions}\label{filter}

For evaluating empirical methods, it is useful to investigate the stability of the observables with respect to the filter functions for our series 
of models. 
In Fig.~\ref{filter} we compare the results computed with the Johnson bandpasses as tabulated by \citet{1990PASP..102.1181B} and the 
CSP-I system response functions.
Qualitatively, the results show the same relations and dependencies. 
The main differences in $(B-V)$ is a systematic shift in $t_{\rm max}(B-V)$ and  bluer $(B-V)$ for the CSP-I system response functions filters at the phase when the Lira 
relation holds. A similar offset in the Lira relation has been observed \citep{2003AJ....125..166K}.
The difference can be understood in terms of the combination of a red and blue shift of the $B$- and $V$-band filters, respectively (see Fig.~\ref{filter}, right).
The smaller wavelength base for the CSP-I response functions compared to the Johnson passbands reduces $(B-V)$ and avoids the strong absorption and 
emission changes at the blue edge of the $B$~band. 
CMAGIC shows the same effect, however its basic features -- the ``knees''-- are minimally affected. 


\subsection{Comparing color-stretch with CMAGIC}\label{CMAGIC}

Both the color-stretch relation and CMAGIC based { methods} are expected to be equally sensitive tools to probe transition objects beyond the 
``$\Delta m_{15}$-cliff''. 
CMAGIC measures the color-brightness space, whereas the color-stretch relation relies on measuring the time of maximum light and the 
reddest $(B-V)$ color. 
CMAGIC draws from rapid changes in the { color-magnitude diagram}.
Color-stretch $s_{BV}$ requires both highly accurate colors and a good time coverage in LCs during the Lira relation.
 In Table \ref{table_mod},
{ we give a range for the theoretical $s_{BV}$ rather than a specific value, 
because our LCs show numerical `jitter' in color on the order of  $\approx 0.015-0.02\unit{mag}$, mostly caused by the radial grid.
The resulting jitter introduces uncertainties of several days in the time { between the maxima in} $B$ and $(B-V)$. }
{ In principle, for the determination of $s_{BV}$, one may use a fitting function to $(B-V)(t)$ around maximum light and the 
slope during the Lira relation to find the maximum in $(B-V)$ using the intercept. However, the former slope is expected to depend on $\rho_c$ and
 $M_{\rm MS}$ (Fig.~\ref{p2t}), and the latter slope is subject to significant uncertainties in the atomic models and the atomic cross sections of forbidden lines.

A possible advantage of CMAGIC based methods compared to a direct use of $(B-V)(t)$ and $s_{BV}$ may be that it, in principle, 
 can be modified  to take into account the shift of the second ``knee'' in $B(B-V)$ since its position depends on $\rho_c$.  
The length of the { second ``leg'' and position of the third ``leg''} may provide one method to characterize $\rho_c$ within DDT  scenarios.

We note that the slope of this second ``leg'' depends on the underlying explosion model. 
For example, mixing processes due to Rayleigh-Taylor instabilities will cause steeper slopes because of increased heating at the 
photosphere and changes of the density structure.
Shells due to pulsations during the explosion will change the colors (see Sect.~\ref{alternatives}).} 
Deviations in the slope may provide physics-based tools to analyze observations.

\section{Benchmarking models}{\label{Analysis}}
\label{successes-and-limitations}

In this section we compare model predictions with observations to gauge the validity and limitations of the models and to demonstrate 
possible applications. 
We use light curve parameters, the time of $B$-band maximum, $t(B)_{\rm max}$, from the template LC fitter {\tt SNooPy} 
\citep{burns11}, and distance moduli and reddening corrections from \citet{burns11}.
We do not optimize the overall fit of the LCs but only align the theoretical time of $t(B)_{\rm max}$ to the observed value. 


\subsection{The brightness decline relation and properties at maximum light}

The comparison of $\Delta m_{15}(V)$ and $(B-V)(t_{\rm max})$ for the forty-three CSP-I SNe~Ia and the theoretical models are shown 
in the right panels of Fig.~\ref{LC_max}. 
We use redshift distances with a Hubble constant $H_0$ of $67\unit{km}\unit{s^{-1}}\unit{Mpc^{-1}}$ \citep{hk96}. 
A combined uncertainty of $600\unit{km}\unit{s^{-1}}$ is assumed due to the peculiar velocity dispersion of galaxy clusters and of 
individual galaxies within clusters ($300-400\unit{km}\unit{s^{-1}}$; \citealt{Bacall96,Masters06,Springob14}) and due to the motion of the 
SN within the host galaxy ($150-300\unit{km}\unit{s^{-1}}$).\footnote{In the case of  SN~2006X, located in  M100/NGC4321, we adopt 
the $\delta$-Cepheii distance of $20.4\pm1.7$(random)$\pm2.4$(systematic)$\unit{Mpc}$ \citep{Mazumdar00}.}
The theoretical models show the same trend as observed in the CSP-I sample and are in good agreement within the uncertainties.


\subsection{Color relations based on SNe~Ia distances}

In this section, we select four SNe~Ia from CSP~I to make detailed comparisons of the shapes of the LCs and color curves between the 
models and observations.
The four SNe~Ia all have low reddening and a rapid-cadence time coverage from well before maximum light to well into the phase of the 
Lira relation.
They also cover a wide range of brightness decline rates from the normal-bright SNe~2005M ($\Delta m_{15,s}(B)=0.96\unit{mag}$), the 
transitional SN~2004eo ($\Delta m_{15,s}(B) =1.39\unit{mag}$) and SN~2005am ($\Delta m_{15,s}(B)=1.49\unit{mag}$), and the 
subluminous SN~2005ke ($\Delta m_{15,s}(B)=1.75\unit{mag}$).
For these four SNe we use distances consistent with the empirical $\Delta m_{15}(V)$ relation given by \citet{phillips99} rather than 
redshift distances, because SNe~Ia with the best coverage at late times are nearby and most affected by peculiar velocities.


The comparison between the theoretical and observed light curves and their properties are given in Figs.~\ref{obs_max} and 
\ref{observations}. 
As expected based on Fig.~\ref{LC_max}, we find that the theoretical $\Delta m_{15}$ relation in the $B$- and $V$-bands are consistent with the observations and well within the uncertainties. 
Our model series can reproduce the $\Delta m_{15}$ relation both in $B$ and $V$ (see Fig.~\ref{obs_max}).
The sizes of the symbols correspond to the { intrinsic} model uncertainties {previously discussed} \citep{h95,HGFS99by02} and become apparent
in the jitter in Figs.~\ref{LC_properties} and \ref{p2t}. 


\subsubsection{Probing the color evolution $(B-V)(t)$}

Qualitatively, all of the $(B-V)(t)$ curves are reproduced by the models with corresponding $\Delta m_{15}$.
The observations show similar color evolution and dependencies (see Fig.~\ref{observations}).
The most obvious success of the models is the qualitative agreement with $(B-V)(t)$. 
This supports our interpretation of the physical origins as discussed in Sect.~\ref{LC}.
Within the range of normal-bright SNe~Ia, similar color evolution at and before maximum light is seen because of the similar ionization 
conditions at the photosphere. 
We consider a range of $\rho_c$ of $0.5-6.\times10^9\unit{g}\unit{cm^{-3}}$ and 
$M_{\rm MS}$ of $1.5-7.\unit{M_\odot}$, which  results in approximately a $0.15\unit{mag}$ dispersion in $(B-V)$ for the phases 
corresponding to the Lira relation.
Subluminous SNe~Ia, those beyond the ``$\Delta m_{15}$-cliff'' such as SN~2005ke, show not only redder color at maximum, but also a 
faster color evolution compared to normal-bright SNe~Ia.
At late times, the $(B-V)(t)$ curves are very similar over the entire range of brightnesses, demonstrating the utility of the Lira relation. 
The most obvious shortcoming of the models is a shift in $t(B-V)_{\rm max}$ by about $3-8\unit{days}$. 
Discrepancies of this order are to be expected (see Sect.~\ref{limits}). 
Therefore, models can guide the differential analyses, but the direct use of $t(B-V)_{\rm max}$ requires calibration by observations.


\subsubsection{Probing a color-stretch relation}

Fig.~\ref{burns} shows the timing of the $(B-V)$ peak versus $\Delta m_{15}$ for the observations and Models 08 -- 25, with additional 
models at lower central densities. 
The functional relationship can be reproduced by the models but $t(B-V)_{\rm max}-t(V)_{\rm max}$ is systematically longer by 
several days. 
Models with lower $\rho_c$ reduce the gap. 
Taken at face value, this may suggest slightly lower $M_{\rm WD}$ and a higher accretion rates than expected for \ce{H}-accreting  
in $\mch$ WD models.
However, the reference times $t_{\rm max}(B-V)$ and $t_{\rm max}(V)$ are set by phases of small change in color and brightness, 
respectively.
As discussed in Sects.~\ref{LC} and \ref{limits}, uncertainties in the fluxes on the order of a few hundredths of a magnitude will produce 
systematic shifts by as large as several days.

The relation shown in Fig.~\ref{burns} is closely related, but not identical, to the color-stretch relation  found by \citet{burns11}. 
\citet{burns11} avoided the problem of shallow extrema altogether by defining the color-stretch $t(B-V)_{\rm max}$ using an analytical 
function to fit $(B-V)(t)$ around maximum light and by data at the time of the Lira relation. 
From our series of models, the secondary parameters of varying progenitor properties impacts $(B-V)$ at late times, and changes in 
$M_{\rm MS}$ affect the shape of the $(B-V)$ evolution around maximum light \citep[see Fig.~\ref{p2t} and][]{2010ApJ...710..444H}. 
Therefore, a secondary parameter, such as $\rho_c$, may introduce a dispersion in the standardization of SN~Ia luminosity using 
the color-stretch relation.
Note that in principle, there may exist a correlation between the secondary parameters and $\Delta m_{15}$ that reduces the dispersion expected from theoretical models. 


\subsubsection{Probing CMAGIC}

Overall, observations agree well with the predictions of CMAGIC with respect to the two ``knees'' and the overall slopes for the rise and 
decline past maximum. 
Moreover, the data show the expected shift of the CMAGIC shapes with SN brightness, which is consistent with the $B$- and $V$-band 
LCs corresponding to their respective $\Delta m_{15}$ (see Fig.~\ref{observations}).   
From CMAGIC, fits are not affected by problems of shallow extrema because we measure the conditions at the photosphere. 

In the case of SN~2004eo at the $\Delta m_{15}$-cliff, the second CMAGIC ``knee'' is bluer by about $0.2\unit{mag}$ than our reference models 
with $\rho_c=2 \times 10^9\unit{g}\unit{cm^{-3}}$. 
As discussed above, this shift may suggest a somewhat lower $\rho_c$ in the framework of $\mch$ models. 
This is consistent with other methods of determining $\rho_c$, specifically with the smaller $\rho_c$ suggested by the late-time IR 
spectra of SN~2005df \citep{tiara15} and the $V$-band  peak-to-tail ratios of SNe~2004eo and 2005M \citep{sadler12,h13}.

Finally, we want to mention some obvious problems.
In the case of SN~2004eo, the initial rise of the theoretical LCs is too fast and its CMAGIC curve seems to be either shifted to the blue by 
$\approx 0.1\unit{mag}$ or it should be brighter by $\approx 0.2\unit{mag}$. 
From the $B$- and $V$-band LCs and the CMAGIC curve, SN~2005ke remains redder by about $0.1-0.3\unit{mag}$ even after the empirical 
$E(B-V)$ correction.
Many of the discrepancies may be attributed to uncertainties in the distances and reddening corrections as discussed by 
\citet{burns11} and in problems of assigning a maximum in the theoretical models.  
However, at least for SN~2005ke, the asymmetry in the spread may hint towards a pulsating delayed detonation model (PDD), as  discussed below in Sect.~\ref{Alternatives}.
{ The normal-bright  SN~2005am and subluminous SN~2005ke have  $\Delta m_{15}$ in the vicinity of values found for Models 23 and 14, respectively.
In Appendix \ref{appendix:spectra}, we show the spectral evolution and discuss the spectral formation in the context of light curves.}


\section{A physics-based approach to data analysis}{\label{Reddening}}
\subsection{A physics-motivated approach to determining distance and reddening}\label{FCR}

After establishing that the model LCs, color curves, and color-magnitude diagrams resemble those of the four well-observed CSP-I 
SNe~Ia, we further examine the validity of the models using forty-three CSP-I SNe~Ia, with good coverage near maximum as the only 
requirement. 
We also take the opportunity to explore physics-motivated methods of determining distance and reddening.


In the previous section, some benefits of the CMAGIC approach were shown and the theoretical color-magnitude curves also provided 
ways to improve upon the technique. 
The incredibly uniform decline in CMAGIC soon past maximum was shown to have a dependence on $\Delta m_{15}$.
The location of the first CMAGIC ``knee'' is sensitive to reddening and maximum brightness, and the location of the second knee 
provides a handle on $\rho_c$ of the progenitor. 
The evolution prior to maximum provides an important, additional anchor that is not degenerate with respect to interstellar reddening.
In practice, to take full advantage of these features in CMAGIC, template color-magnitude curves should be constructed using 
well-observed SNe~Ia, perhaps with guidance from the models to disentangle effects due to secondary parameters.
Because the templates depend on the brightness, $\Delta m_{15,s}$ needs to be used to select the correct template. 
Since such templates are not yet available, we opted to use the modeled LCs and color curves directly as an exploratory test first.

Our Models 08 -- 27 span the brightness range of observed SNe~Ia (see Fig. 3). 
We then interpolate between the models to produce a grid of $B$- and $V$-band LCs and color-magnitude curves with respect to the 
decline rate of the model light curve. 
{ Here, we neglect the variations in the progenitor parameters, as these are most relevant at phases where most of our forty-three 
SNe~Ia do not have coverage, and which have not yet been included in the empirical methods used as a benchmark.} 

The grid of models are fit to the data via $\chi^2$ minimization using five parameters: 
\begin{enumerate*}
\item[1)] decline-rate parameter, 
\item[2)] time of maximum relative to the explosion,
\item[3)] distance modulus, 
\item[4)] color excess, $E(B-V)$, and
\item[5)] total-to-selective extinction ratio, $R_V$. 
\end{enumerate*}
The fitting procedure is a two step process as described below:

{\sl Step 1:} 
Fitting the model to $B$- and $V$-band LCs simultaneously provides a global decline-rate parameter $\Delta m_{15,s}$ and a time of 
maximum relative to the explosion. 

{\sl Step 2:}
Based on the $\Delta m_{15,s}$ from the LC fit, a model color-magnitude template is selected.
It is then fit to the data of each SN~Ia in the CMAGIC color-magnitude space. 
The fitting parameters here are distance modulus, $E(B-V)$, and $R_V$.

\subsection{Test of the fitting procedure}

In the following, we will discuss three consistency checks for the models: 
\begin{enumerate*}
\item[1)] the color at maximum light, 
\item[2)] the distance modulus, and 
\item[3)] the reddening. 
\end{enumerate*}
The first check is an internal test for the quality of the theoretical templates, whereas  the latter checks probe the accuracy of the derived 
and absolute quantities, such as the reddening and the absolute brightness.

{\sl 1)}
One consistency check of the models is obtained by comparing the theoretical color at maximum light, $(B-V)_0$, with the intrinsic color 
expected by the parameter fit, which depends on global properties (Column 2 of Table~\ref{table_fit}).   
The colors are consistent within an average of $\approx0.05\unit{mag}$, ranging from $0.008\unit{mag}$ (SN~2005ki) to 
$0.12\unit{mag}$ (SN~2007ai).
Distant SNe~Ia tend to show larger discrepancies than the average, for example SNe~2006gt and 2007ai.  
Errors of this size can be expected from models and may also be partially attributed to observations. 
This test is non-trivial because $\chi^2$ is minimized to global quantities versus the color at maximum light.
For example, the corresponding $\chi ^2$ are of the order { $0.6$} for sub-$\mch$ models discussed in Sect.~\ref{alternatives}.  
{\sl 2)}
For all SNe~Ia, the resulting distance modulus, $\mu=m-M$, is well within the range given by other methods (Column 5 of 
Table~\ref{table_fit}).
The agreement with respect to redshift distances is very good for distant SNe~Ia, but may be as large as $0.4-1.0\unit{mag}$ for local 
SNe~Ia such as SNe~2007on, 2007af, and 2005am (see Fig.~\ref{dist}). 
Two SNe, 2007on and 2011iv, are slightly subluminous with $\Delta m_{15,s}(B)$ of $2.011$ and $1.72\unit{mag}$ and 
$\Delta m_{15,s}(V)$ of $1.110$ and $1.07\unit{mag}$, respectively.
These two SNe are in the same same host, NGC 1404, and have low extinction (Table~\ref{table_fit}). 
The two values of $\mu$ differ by $0.15\unit{mag}$, which is one indicator for our combined error in distances and reddening.

An alternative test is SN~2006X, a highly reddened SNe~Ia in M100 with a $\delta$-Cepheii based distance. 
Our  $\mu($SN~2006X$)=31.7\unit{mag}$ is consistent with $\mu($M100$)=31.5\pm0.2$(random)$\pm0.3$(systematic)$\unit{mag}$
\citep{Mazumdar00}.  

{\sl 3)}
Reddening results in a shift of the CMAGIC curve both in color and brightness.
To first order, the reddening by the ISM can be estimated by using a color-brightness relation at maximum light or assuming the same 
intrinsic colors making use of the Lira relation \citep{lira95} and assuming that $R_V$ and $R_B$ are known. 

Although long suspected, observational evidence has shown that extinction laws are not the same in our Galaxy when compared to other 
galaxies \citep{1993AJ....105..301L}. 
Multi-color fits allow information to be added in order to address the question of extinction.
{ Using a wide range of filters in the optical and NIR,  \citet{2003AJ....125..166K}, \citet{burns11} and \citet{2014ApJ...797...75M}
showed that the  average $R_V$ in other galaxies is closer to $2.8 \pm 0.4 $ and that 
highly reddened objects give $R_V \approx 1.7$. Although the individual spread is large, the average values are systematically smaller
than the $R_V=3.1$ commonly adopted for the Milky Way galaxy.} 
Making use of multi-color LC fits and template LCs, \citet{burns11} confirmed that measured values of $R_V$ tend to be lower than the 
Galactic value with a wide spread between $0.8-4.3$ and individual uncertainties span a significant range of the observed $R_V$. 

As a benchmark, we use the empirical reddening corrections from {\tt SNooPY} 
\citep{burns11,2014ApJ...789...32B,2015ascl.soft05023B} based on the reddening law of \citet{1999PASP..111...63F}.
A detailed discussion is beyond the scope of this paper.
We pick as examples the highly-reddened SNe~2006X, 2007S, and 2007le. 
{ Our best fits of the empirical values by \citet{burns11} for  $(E(B-V),R_B)$ are:
\begin{itemize}[nosep]
\item $(1.182,2.4)$ compared to $(1.36\pm 0.026, 1.8 \pm 0.1)$ for SN~2006X, 
\item $(0.517, 2.9)$ compared to $(0.478 \pm 0.026, 1.9 \pm 0.2)$ for SN~2007le, and
\item $(0.396,2.7)$ compared to $(0.388\pm0.023,1.6 \pm 0.2)$ for SN~2007le.
\end{itemize}
If we use a linear expansion for  $\Delta R_{B,V} \approx R_{B,V}/E(B-V)\times(\Delta(m-M)+\Delta E(B-V))$, our errors in $R_B$ are 
$\approx 0.5-0.8$. }

Overall, the values for the reddening agree but our $\Delta m_{15,s}$-CMAGIC based values for $R_B$ are somewhat larger, possibly due to systematics in the models. 
In Fig.~\ref{dm15r}, we show the loci of all forty-three CSP-I objects with coverage of maximum light and { redshift-based distances}.
The importance of reddening corrections is apparent, but overall, our values compare with the empirical method.

%
%
%
%

\section{Confronting theory with observations for the full CSP-I SNe~Ia sample}{\label{Analysis2}}
After establishing that the theoretical models resemble observations for the limited set of SNe presented in 
Sect.~\ref{successes-and-limitations}, we will apply the results to the broader CSP-I data set. 
The goal is to study the spread around the shape of the theoretical predictions and their shift.
The observed LC parameters have been optimized with respect to distance, time of explosion, and extinction, and we interpolate between 
the grid of models. 
In Sect.~\ref{Reddening}, the $\Delta m_{15,s}$ relations in both the $B$- and $V$-bands have already been employed as a consistency 
check for our fitting parameters of the forty-three CSP-I SNe~Ia.
{ Because we use the brightness-color relations, CMAGIC, and theoretical templates, the resulting 
fits to the $\Delta m_{15}$ relation will improve 
by construction, however these fits will not provide further constraints.} 
Our discussion will focus on the twenty-one CSP-I SNe~Ia which allow us to study the color relations (see Tables~\ref{table_obs} 
and \ref{table_fit}). 

We grouped the observations in a narrow range of $\Delta m_{15,s}\approx 0.6\pm 0.1 $, referred to as Series 1, and the entire range of $\Delta m_{15,s}$, 
referred to as Series 2. 
The observed and theoretical $B$- and $V$-band LCs, $B(B-V)$, $V(B-V)$, and $(B-V)(t)$ are shown in Figs.~\ref{Series_2} and 
\ref{Series_1}.

All of the Series 1 SNe~Ia are tightly clustered both on the rising and declining ``legs'' of the CMAGIC curve, with a width of 
$\approx 0.1\unit{mag}$, despite a dispersion in the monochromatic LCs of $\approx 0.5\unit{mag}$. 
The dispersion both during the Lira relation phase in $(B-V)(t)$ and on the third ``leg'' in the color-magnitude diagram are about $0.25\unit{mag}$, which can be 
attributed to variations in $\rho_c$ { and can explain both the shift in brightness of the third ``leg'' and the bluer color at the 
transition from the second to the third ``leg'', discussed in Sect. 3.2.}

If this same consistency were seen for a larger sample of SNe~Ia, it may justify the use of the theoretical templates for $\Delta m_{15,s}$ 
and CMAGIC. From the models, we would expect a small dispersion in the LC parameters within a homogeneous class of explosion scenarios.
One possible exception is SN~2008fp, which shows an overall flatter $(B-V)(t)$ evolution than the other objects in the Series.
However, this object also lacks pre-maximum colors and it shows a small $\Delta m_{B}$ compared to $\Delta m_{V}$ { (see Table 1).}

Series 2 has a wide spread in $\Delta m_{15,s}$ and spans most of the range for the bright end of SNe~Ia beyond the 
$\Delta m_{15}$-cliff and confirms the continuous transition between normal-bright and subluminous SNe~Ia. 
The existence of these correlations without a ``break'' between normal-bright and subluminous SNe~Ia may suggest the same dominant 
explosion scenario. 
As in Series 1, the dispersion in CMAGIC during the Lira relation (\textit{Phase 4}) and the curve offset is consistent with variations in 
$\rho_c$. 
In particular, SN~2005iq is an example of a transitional object between normal-bright and subluminous SNe~Ia that is bluer during the Lira 
relation (see Fig.~\ref{Series_1}, lower panel) and rather similar to SN~2007on, studied in detail by Gall et al. (2017, submitted to A\&A).

\section{Alternative explosion scenarios}{\label{Alternatives}}
\label{alternatives}


Finally, we  put our results into the context of different scenarios for thermonuclear explosions. 
A comprehensive test of the full variety of explosion scenarios and their parameter space does not exist. 
Nuclear physics governs the structure of the WD, the specific energy release, and burning products given by quasi-nuclear equilibrium of 
explosive \ce{C}-burning, incomplete \ce{O}-burning, and full NSE.
Due to the ``stellar amnesia'' of thermonuclear explosions of WD, to first order most models show LCs rising on time-scale of 
$\approx 15-25\unit{days}$ followed by a phase dominated by the radioactive decay of \ce{{}^{56}}Co and maximum spectra dominated by 
incomplete \ce{O}-burning and later on by iron-group elements. 
Therefore, our goal is to discuss the general properties of the scenarios and to show using several examples that color relations provide a 
sensitive tool to probe the diversity of SNe~Ia progenitors.

The consensus picture is  SNe~Ia result from degenerate carbon-oxygen WDs that undergo a thermonuclear runaway \citep{hf60}, and that they originate from close binary stellar systems.
{\sl Potential progenitor systems} may either consist of two WDs, a so-called double degenerate system (DD), and/or a single WD and a 
main sequence, helium or Red Giant star, so called single degenerate systems (SD).
Within the general picture for progenitors, four classes of {\sl explosion scenarios} are discussed that are distinguished by the triggering 
mechanism of the explosion:

{\sl 1)} 
Dynamical or violent merging of two carbon-oxygen WDs on time scales of seconds where the explosion is triggered by heat during the 
merging process \citep{iben,webbink84,benz90,loren09,WCMH09,isern11,pakmor10,2012ApJ...749...25G}. 
The explosions will produce a wide variety of envelope masses ranging from well below $\mch$ to super-$\mch$ and very asymmetric 
density distributions in their ejecta.
This class of objects will show a significant continuum polarization that has not be observed, and we must expect a directional 
dependence of the luminosity, which would result in a wide spread of observed $\Delta m_{15}$ and $\Delta m_{15,s}$
\citep{2006NewAR..50..470H,patat12,2016MNRAS.455.1060B}.

{\sl 2)} 
Explosions of sub-$\mch$ carbon-oxygen WDs triggered by detonating the surface \ce{He} layer originating from a helium star or a 
low-mass WD companion \citep{wwt80,n82,livne1990,hk96,kromer2010,pakmor2012,Diehl2015}. 
Overall, this class of models produces only small deviations from symmetry. 
In most simulations, the resulting envelope has a sizable amount of \ce{{}^{56}Ni}, which has been invoked to understand observations of  
early \ce{{}^{56}Ni} $\gamma$-ray lines in SN~2014J \citep{Diehl2015}. 
Because the specific explosion energy does not depend on $M_{\rm WD}$, we must expect $(B-V)(t)$ evolutions with a wide dispersion 
independent of brightness. 
The corresponding models in \citet{hk96} show no ``$\Delta m_{15}$-cliff'' and no tight $\Delta m_{15,s}$ relation.
In Fig.~\ref{He_properties}, LC properties are shown for the double-detonation (or \ce{He}-detonation) subluminous Model HeD6 and the 
normal-bright Model HeD10 with masses of $0.78 $ and $1.02\unit{M_{\odot}}$. 
These models produce $0.25$ and $0.52\unit{M_{\odot}}$ of \ce{{}^{56}Ni} and show peak brightnesses of $M_V=-18.5$ and 
$-19.6\unit{mag}$, respectively \citep{hk96}. 
The color evolution is characterized by blue colors immediately after the explosion and a slow color evolution after maximum.  
The early blue color is caused by a small amount of radioactive \ce{{}^{56}Ni}, and blue colors after maximum are caused by the low mass 
and high expansion velocities, which result in rapidly-receding photospheres and energy input above the photosphere soon after 
maximum light. We note that the subluminous model has a significant amount of intermediate mass elements. 
As an inefficient heater, after the outer \ce{{}^{56}Ni} becomes optically thin, Model HeD6 shows a similar evolution in CMAGIC compared to our 
DDT models (see Fig.~\ref{He_properties}). { For the same maximum brightnesses,}  
sub-$\mch$ models tend to show faster rising LCs. { For our normal-bright HeD model, a significantly larger $\Delta m_{15}$ is seen
with a steep dependence on brightness, similar to $\mch$ models beyond the ``$\Delta m_{15}$-cliff''.}
Recently, some evidence of $\gamma $-rays from \ce{{}^{56}Ni}  have been reported, although it is not clear whether INTEGRAL has seen 
narrow or very wide lines of $1,600$ or $\approx 30{,}000\unit{km}\unit{s^{-1}}$ \citep{Diehl2015,Isern2016}. 
However, early blue colors are not observed in the CSP-I sample and, if observed a few days after the explosion, they may be caused by 
interaction rather than radioactive heating.

We note that this class of models has not been fully explored.
The blue color may be avoided by increasing the mass of the carbon-oxygen core because the required helium shell size decreases, 
which can then trigger a central detonation \citep{nomoto82,woosley94,hk96}.
This path has been studied by, e.g., \citet{kromer2010}, \citet{Shen2014} and \citet{pakmor2016}, who published a series of \ce{He}-shell triggered 
sub-$\mch$ models with little \ce{{}^{56}Ni}.
In theory, models even without \ce{{}^{56}Ni} cannot be excluded. 
However, their models show $\Delta m_{15,s}$ in $B$ and $V$ to be too large by a factor of two and, for normal-bright SNe~Ia, the 
$(B-V)$ color is too red by about one magnitude at all phases. 
 
{\sl 3)}
Explosions of carbon-oxygen WDs with a mass close to $\mch$, as discussed in this paper. 
A variation from the classical DDT scenario are models in which the WD stays bound after the initial deflagration burning phase, and 
subsequently, undergoes a pulsation prior to the deflagration to detonation transition \citep{kmh93,hkw95,hk96,q07,dessart14}. 

{ 
In PDD (pulsating delayed detonation) models, nuclear burning prior to the detonation phase will reduce the binding energy of the WD. Variable  amounts 
of burning during the deflagration phase  will produce a continuum from a strongly bound to an almost unbound WD with a smooth 
transition to classical DDT models. 
Moreover, multiple pulsations will change the amplitudes of the PDDs and, thus, the amount of unburned carbon and oxygen, and 
mixing during the pulsations can be expected to change the composition of the shell.
In all current simulations, the detonation is triggered by mixing of burned and unburned material through the  so-called Zeldovich mechanism.
It may occur anywhere during the contraction and expansion phase of the WD, and models differ by the amount of mixing.} 
The PDD models introduce a large diversity into the empirical relations. 
Depending on mass and amplitude of the pulsation, { our PDD models with large amplitudes} produce a longer pre-heating phase 
with red colors at maximum light, correlated with broad maxima in $(B-V)(t)$, broader $B(B-V)$ and outer layer beyond the shell
of faster expanding C/O-rich layers. 
Pulsations change the density structure, producing longer diffusion time scales and redder colors, which will impact the second leg of the 
CMAGIC curve. It is likely that PDDs do contribute to the SNe~Ia population. 
In our sample, the subluminous SN~2005ke may be a candidate. 
PDDs have also been suggested for some of the brightest SNe~Ia with slow decline rates, { red colors}, and large $s_{BV}$ and plateaus in the velocity 
evolution, e.g.,  SNe~1990N and 1991T \citep{hk96,quimby06}.\footnote{{ The diversity of PDDs may also be exemplified by a recent study by
\citet{dessart14} who find for their models colors are similar to DDT models at maximum light and significantly bluer at early times.}}

\section{Final Discussion}\label{DiscussionConclusions}

In Sect.~\ref{Analysis}, we tested the theoretical predictions of our baseline models with four supernovae with low reddening
from the CSP-I survey, i.e., SNe~2004eo, 2005M, 2005am, and 2005ke. 
We find good overall agreement over the brightness range and discuss their limitations while using empirical values for distance and  
reddening.

In Sect.~\ref{Methods}, we linked observables to the underlying explosion physics for our baseline scenario. 
Colors probe the conditions at the photosphere, and absolute fluxes are governed by the energy input and transport effects at deeper 
layers.    
Opacities are dominated by lines and electron scattering, and to first order, the average opacities increase with heating.
However, their distribution in frequency space dominates the color evolution.
The lines act as ``absorption rods'' in front of a bright photosphere early on and as ``emission rods'' at later phases.
The empirical relations are all based on the same physical underpinnings but have different virtues as diagnostic tools depending
on their use of flux and color information. 

Key ingredients are the amount and distribution of \ce{{}^{56}Ni}, an intense heating phase lasting $1-2\unit{weeks}$ from \ce{{}^{56}Ni} 
decay, the strong drop of the opacities at low temperatures, a larger contribution of many weak lines in the $B$ band compared to the $V$ band, and 
subsequently, the increasing energy released above $R_{\rm phot}$ coupled with the photosphere entering layers with shallow density 
gradients. { As discussed in the Introduction and Sect. 3, $\Delta m_{15}$ can be understood by the stored energy and the diffusion time scales.}
Whereas $\Delta m_{15}$ is dominated by layers below the photosphere, the color relations are dominated by the similarity in density 
and abundances at the photosphere during similar phases for models ranging from normal-bright to subluminous, SN~1991bg-like, 
objects. 

During the first $1-2\unit{weeks}$, the {\sl heating phase}, adiabatic cooling and re-heating of the envelope by the intense 
energy release of \ce{{}^{56}Ni} dominate, with a half-life $t_{\rm decay}=6.1\unit{days}$ and an increasing propagation 
of $\gamma$-rays. 
Early on, the outer layers without \ce{{}^{56}Ni} will be partially transparent because adiabatic cooling and the lack of \ce{{}^{56}Ni} 
causes low temperatures and, thus, low opacities. 
This effect may lead to photospheric velocities seen in spectra that increase during the first hours to a few days for normal-bright and 
SN~1991bg like objects, respectively.
In the {\sl photospheric phase} we find two distinct regimes at work for normal and subluminous SNe~Ia. 
In subluminous SNe~Ia, less \ce{{}^{56}Ni} means less overall brightness and faster reddening, the origin of the $(B-V)(t)$ relation. 
For normal-bright SNe~Ia, the minimum in $(B-V)$ corresponds to recombination at the photosphere from doubly- to singly-ionized 
metal. 
For both, the reddening in $(B-V)$ is a result of the photosphere being formed in the envelope where there are shallow density profiles. 
However, for subluminous SNe~Ia, this phase is entered earlier because of the faster-receding photosphere.  
By the time of the {\sl semi-transparent phase} the photosphere has receded to the \ce{Ni} region with a similarly shallow density profile. 
The similarity of the conditions causes the Lira relation and its stability.

In Sects.~\ref{limits} and \ref{filter}, we discussed the limitations of our models from the perspective of stability and the underlying 
physics, namely radiation transport and uncertainties in atomic data. 
As a consequence, small variations introduce significant uncertainties in the exact times of maximum brightness, the Lira relation, and 
color-stretch. 
We showed that $(B-V)(t)$ depends sensitively on the filter functions and exposes model and observational uncertainties.  
For high-precision cosmology, in the context of purely LC-based methods, this again emphasizes the importance of consistent data sets, 
and that the accuracy may be limited by $K$-corrections. 
The problems may call for a large number of SNe~Ia in narrow red-shift bins or the inclusion of spectra.

From theory, all of the empirical relations are subject to secondary parameters introduced by specifics of the progenitor, because among 
other considerations, the specific nuclear energy available will depend significantly on $M_{\rm MS}$ and potentially on $Z$ 
\citep{timmes03,hwt98}.
Within spherical DDT scenarios, \ce{{}^{56}Ni} is produced both during a deflagration and a detonation phase, which 
dominates \ce{{}^{56}Ni} production for normal-bright SNe~Ia. 
In $\mch$ WD models, the amount of \ce{{}^{56}Ni} production is dominated by the parameter $\rho_{\rm tr}$, because most of the 
\ce{{}^{56}Ni} is produced during the detonation phase of burning. 
However, in subluminous SNe~Ia, a variation in $\rho_c$ results in large increases in the spread of $\Delta m_{15}$. 
We showed that the dispersion grows from about $0.1$ to $0.7\unit{mag}$ { between}  normal-bright and very subluminous SNe~Ia, respectively.
In subluminous SNe~Ia, lower opacities result in a significant contribution of energy input from the central \ce{{}^{56}Ni} layers already by 
maximum light. 
For both normal and subluminous SNe~Ia, our models showed that variations in the progenitor, namely $\rho_c$, will result in a shift of 
up to $0.2\unit{mag}$ of the Lira relation. 
For normal-bright SNe~Ia, $\rho_c$ has an effect on LCs some $40-50\unit{days}$ after the explosion but may elude detections around 
maximum light because of the similarity of both the spectra and colors.

Note that a method for identifying spectroscopic ``twins'' during the photospheric phase has recently been proposed by 
\citet{2015ApJ...815...58F} to improve distance measurements { and scatter in the Hubble diagram} within $0.083\pm0.012\unit{mag}$ and $0.072\pm0.010\unit{mag}$ for SNe~Ia in the 
Hubble flow..
This spread is smaller than expected by a factor of $2$ for normal-bright SNe and by a factor of $10$ for 1991bg-like SNe~Ia, even in the 
framework of one specific scenario. 
An answer is beyond the scope of this paper, but several possibilities may include: 
\begin{enumerate*}
\item[a)] selection effects in the observations due to a limited brightness range, and that the spread is not indicative of systematic shifts 
with redshift, or 
\item[b)]  there exist intrinsic correlations between brightness and the underlying progenitors, as discussed in Sect.~\ref{Methods}.
\end{enumerate*}

CMAGIC { and brightness-color based methods}, and differential analyses of light curves are resilient to variations in secondary parameters, and the latter does not depend on 
reddening corrections. 
We showed that the resilience of the {\sl first two methods} is a result of the compression of slow time evolutions and emphasizes the transition between 
different regimes.\footnote{Differential light curve analyses in $U$-, $B$-, and $V$-band may have higher accuracy in detecting intrinsic 
variations, providing a direct { link} to the progenitor with $\mch$ explosions, and they do not depend on interstellar reddening 
\citep{2010ApJ...710..444H,sadler12,h13}.}
However, the theoretical CMAGIC templates have a systematic shift with brightness or $\Delta m_{15,s}$.
We propose a combination of $\Delta m_{15,s}$ and CMAGIC to measure the properties of ISM dust, namely $E(B-V)$ and 
$R_B$, and it is sensitive to changes in the explosion scenario because the two long linear ''legs`` on the rise and decline provide 
stability. 

In Sect.~\ref{Reddening}, we presented a $\Delta m_{15,s}$, { brightness-color}, and template-based approach for both the $B$- and $V$-bands
to recover distances and ISM reddening laws using interpolated theoretical templates as a proof of principle, and to be able
to perform a detailed analysis of a large set of SNe~Ia.  
For verification of the theoretical relations and their absolute calibration, we used twenty-one CSP-I SNe~Ia with good coverage on both 
the rise and decline from maximum. 
The maximum colors are consistent within $\approx 0.05\unit{mag}$. 
Within the error bars, our distance for the highly reddened SN~2006X is consistent with the $\delta $-Cepheii distance. 
The same distance, within the error bars, are found for SNe~2007on and 2011iv, which are in the same host galaxy (see Gall et al. 2017, 
submitted to A\&A). 
The reddening corrections inferred from CMAGIC analysis,  $E(B-V)$ and $R_B$, are consistent with those obtained by empirical well-calibrated methods (Fig. \ref{dm15r}).

For the same CSP-I sample, the color evolution, $B(B-V)$ and $V(B-V)$  show a small dispersion for normal-bright SNe~Ia and a strong,
systematic shift towards subluminous objects, as described in Sect.~\ref{Analysis2}.
We showed that $\mch$ models and DDT models provide an explanation of the observed empirical relations and their 
properties, and we submit that they can guide the discovery of secondary parameters with which to probe the diversity of SNe~Ia. 
Observations are needed to solve the conundrum of why we see evidence for deflagration burning while at the same time 
requiring the expected extensive mixing by Rayleigh-Taylor instabilities to be suppressed. 
Realistic predictions of mixing would improve the use of SNe~Ia for cosmology { because variable mixing causes a shift in the individual 
$\Delta m_{15}$ (Fig. \ref{LC_properties}), which may lead to a spread or systematic shift with redshift in the luminosity decline rate relation.}
We discussed the very early phase of re-heating, which may result in an increase in the velocity of material at the photosphere
lasting between hours and several days as a probe for mixing processes.

From Fig.~\ref{Series_1}, we see that SN~2005ke becomes too red well after maximum light and the CMAGIC relation is too wide,
even after corrections for reddening. 
This may indicate a continuum of explosions ranging from classical DDT models to PDD, in which the WD stays bound to a various 
degree, { namely lifted in its gravitational potential during the pulsation},
 and may even undergo multiple pulsations with various amplitudes as discussed at the end of the previous section.
In this framework, classical DDT models are very similar to low-amplitude, single PDDs. 
Note the very subluminous SN~1999by can be reproduced with classical DDT models. 
We may have a continuum of DDT/PDDs for a range $\Delta m_{15}$, as suggested by \citet{2015A&A...573A...2S}
  
In Sect.~\ref{Alternatives}, we showed that the color relations are not generic for  most of the explosion scenarios, and that
the differences can be understood within the theoretical framework. 
Whereas some $\Delta m_{15}$ relation must be expected for most explosion scenarios because of the effects of opacity,  the color 
evolution adds the physical condition at the photosphere as another constraint.  
{ For example, both sub-$\mch$ and a DDT may have the same $\Delta m_{15}$ but different $(B-V)(t) $ and $B(B-V)$.}
We discussed the color relations for a wide range of scenarios, including normal-bright models, subluminous \ce{He}-triggered 
sub-$\mch$ models, and several sub-$\mch$ models from literature.   
If sub-$\mch$ explosions are a dominant scenario, they must have high masses in order to reduce the amount of \ce{He} needed to 
trigger the explosion { \citep{nomoto82,woosley94,hk96}, and they should have structures very similar to $\mch$ models \citep{hk96,Shen2014}.}

{ The continuous light curve relations in the observations may suggest that one scenario dominates over a wide range of $\Delta m_{15}$.
 A change of slope in the luminosity decline rate relation does not imply that we must have different progenitor masses at different decline rates 
as suggest by \citet{Scalzo2014}. Though classical delayed-detonations and pulsating delayed detonations are good candidates for the
dominant class, SN1991T-like objects and two recently identified new subclasses, SNe~Iax  \citep{li02cx03,li11} and super-Chandrasekhar SNe Ia \citep{Howell2006}, 
demonstrate the diversity of thermonuclear explosions.} 

Here, we want to address the potential use of other colors.
Broadband $(B-V)$ color is suitable because it is based on variations in the temperature and the ionization balance at the photosphere. 
{ Based on detailed comparisons of synthetic and observed spectra for normal bright and subluminous SNe~Ia \citep{1998ApJ...496..908W,HGFS99by02,tiara15}, 
we can expect that the $(V-H)$ color will add new information because of the time-dependent broad emission features by iron-group elements in $H$ and their velocity shifts. 
If the $\mch$ explosion is the dominant channel, 
$(V-H)$ may be expected to produce stable relations because the $H$-band is 
dominated by iron-line emissions in the $1-2\unit{weeks}$ after maximum.}
The $U$-band is an excellent indicator for the metallicity, velocity gradients, and ongoing interaction of the envelope with the surroundings,
{ however it requires detailed fits of LCs, flux, polarization spectra, and line profiles to separate the various effects and is notoriously difficult to calibrate.} 
Colors involving $R$, $I$, $J$, and the IR can be expected to be prone to systematics because of a more complicated color evolution 
that depends sensitively on single emission lines, such as the \ion{Ca}{2} IR triplet or the emission of iron group lines contributing at the 
edge of the $J$-band. 

\section{Conclusions}\label{Conclusions}
We presented a study of optical LC properties and their physical underpinnings using twenty-six classical DDT  
 models as a baseline, in combination with modern photometry, namely well-observed CSP-I SNe~Ia  described in 
 Sect.~\ref{Observations}. 
Alternative scenarios have also been addressed, including explosions of sub-$\mch$ WDs. 
We revisit the brightness-decline relation $\Delta m_{15}$, the Lira relation where the $(B-V)$ color at $40-60\unit{days}$ { past explosion}
 is similar for all 
SNe~Ia, and present a study of the color-stretch relation $s_{BV}$, the temporal color evolution $(B-V)(t)$, and the origin of the CMAGIC 
relation $B(B-V)$. 
We showed that an understanding of empirical relations allows us to evaluate new pathways for the analysis of observations,
and for high-precision cosmology, to evaluate their stability with respect to uncertainties in the observables and population shifts intrinsic 
to progenitors, explosion scenarios, and the ISM properties. 

{ The observed LC and color curves show continuous relations from normal-bright to subluminous SNe~Ia}.  
The models naturally produce the entire observed range of luminosities and color properties, advocating for a single explosion mechanism 
from normal-bright to subluminous SNe~Ia as a dominant channel with a likely contribution from PDD models.
The models even reproduce the subtle slope change for fast-declining SNe~Ia in the peak luminosity versus decline rate plot, which has 
sometimes been interpreted as a different explosion mechanism \citep{2017arXiv170206585D}. 
We cannot rule out other explosion scenarios { as dominant contributors to SNe Ia}, but suggest that the final outcome 
should be similar to $\mch$ models. 

The models we use have not been tuned to reproduce the correlations seen. 
In fact, most of the explosion models have been taken from previously-published works and have only been supplemented to include 
progenitor variations for subluminous SNe~Ia. 
The sub-$\mch$ models investigated here could not reproduce the correlations that are expected from observations.

From the theory side, the color-stretch parameter $s_{BV}$ is obtained from the intercept of two portions of the $(B-V)$ color curve that 
are dominated by different physical processes and thus governed by different physical parameters. 
Correction effects from secondary parameters, such as $\rho_c$, affect the different phases following the explosion to a varied 
degree, so use of $s_{BV}$ may obscure underlying physical effects.
The motivation for using the color-stretch parameter is that it sorts fast-declining SNe Ia much better than $\Delta m_{15}$  
{ because,  for low luminosity SNe~Ia, the inflection in optical colors occurs before 15 days.} 
is that $15\unit{days}$ runs into the inflection point of the light curve for fast 
decliners. 
We suggest that using stretch-corrected decline rate measurements, $\Delta m_{15,s}$ is another option that circumvents the problem.

We find great potential in the combination of CMAGIC and $\Delta m_{15,s}$ to determine reddening and distance.
Comparisons of predictions to the observed CMAGIC curves will help eliminate models.
The intrinsic SN  colors characterize the decoupling region in the ejecta and the overall brightness measures the diffusion time scales.
The small dispersion in the CMAGIC curves for a given $\Delta m_{15,s}$ also point to a dominant scenario for progenitors and 
explosions.
The Lira relation works at the $0.2\unit{mag}$ level, as that portion of the color curve is influenced by variations in central density.
(see Sect. 3 and Fig. 6).
Besides decline rate and intrinsic color, central density has the largest influence on peak magnitude, especially for subluminous and fast-
declining SNe, and is key to improve SN~Ia precision for the next generation of dark energy experiments. 
Constraints on central density can only be obtained at later times, some 40 to 60 days past maximum, and present a challenge for 
high-redshift studies. 

Finally, we want to discuss the limitations of this study. 
The CSP-I sample broke new   ground in terms of consistency and accuracy, but the actual number of objects is limited 
and we cannot exclude many of the possible explosion scenarios. 
Moreover, we cannot exclude that explosions of other than current $\mch$ WD scenarios may equally well reproduce the observations.
Physics-based semi-empirical methods are a promising path to separate effects from the ISM and the distance, but it may take empirical 
templates for CMAGIC for use in high-precision cosmology.
Spherical DDT  models allow for a good approximation of the density and abundance structures observed in SNe~Ia, but they are 
obviously only an approximation. 
Within this scenario, off-center DDTs are to be expected and some evidence has been seen in SN-remnants, early time 
spectropolarimetry, and late time spectra \citep{fesen07,2006NewAR..50..470H,patat12}.
, all of which unfortunately will not be available for high-redshift SNe~Ia. 
The continuum of DDT to PDD models needs to be explored in a systematic way, and we must understand how to suppress 
Rayleigh-Taylor instabilities, for example by invoking magnetic fields. 
We found models with different brightness but similar conditions at the photosphere that will appear as ``spectroscopic twins'' \citet{2015ApJ...815...58F}. 
A study of how similar the spectra are will be addressed in a forthcoming paper.
For all scenarios, we must expect variations in $Z$ and $M_{\rm MS}$, amongst others, will influence the explosion energy. 
Moreover, for most scenarios, the parameter space has yet to be explored to answer whether one can find solutions which 
closely resemble $\mch$ models in their basic properties of being spherical with a layered chemical structure. 
If the WD masses are sufficiently high so that electron capture elements are produced, we need LCs or spectra in the $1-2\unit{months}$ 
after maximum light despite the focus on earlier times for many current and upcoming SN surveys. 

\acknowledgments

We thank the referee for useful suggestions that improved the presentation of our results.  
P. Hoeflich acknowledges support by National Science Foundation (NSF) grants AST-0708855, and AST-1008962. M. Stritzinger acknowledges generous support from the Danish Agency for Science and
Technology and Innovation realized through by a Sapere Aude Level 2 grant, and also to a research grant (13261) from the VILLUM FONDEN.
The CSP is supported by NSF grants AST-0306969, AST-067438, AST-1008343, AST-1613426, AST-1613455, and AST-1613472.  
D.J.S. acknowledges support from NSF grant AST-1517649.
T. Diamond acknowledges support through an appointment to the NASA
Postdoctoral Program at the NASA Goddard Space
Flight Center, administered by Universities Space Research
Association under contract with NASA.

\begin{appendix}
\section{Spectral Formation in Context of Light Curves}\label{appendix:spectra}

{ 
 As mentioned above, spectra of SNe~Ia can be understood by overlapping P-Cygni profiles on top of a quasi-continuum.
 To first order, the spectral sequence and expansion velocities can be understood in terms of a receding photosphere which exposes
layers  of C/O, explosive carbon, oxygen, incomplete Si-burning, and NSE in the inner region. Nuclear physics rules.
 However, the time-sequence and spectral details will depend on the explosion physics, the resulting structure, and equally important, 
 uncertainties in the atomic physics which affect the light curves to a much smaller amount.  
 For illustration and to emphasize the latter, we show the time sequence for the normal bright SN~2005am (Fig.~\ref{SN2005am}) and  the subluminous SN~2005ke (Fig. \ref{SN2005ke})
because their $\Delta m_{15}$ are, by chance, in the vicinity of values found for our Models 23 and 14, respectively (Fig. \ref{observations}). 
The comparison does not constitute spectral fits or analyzes. For dedicated fits and identification of features, we will refer 
to previous papers mentioned above \citep[e.g.,][]{h95,1998ApJ...496..908W,HGFS99by02,patat12}. We will focus on spectral properties 
relevant for the LCs and the color evolution.

The observations have been taken from the Weizmann Interactive Supernova data REPository  \citep[WISeREP;][]{2012PASP..124..668Y}. The SNe~Ia data
were obtained using the Las Campanas Observatory du Pont Telescope in Chile, the FLWO Telescope on Mt. Hopkins in Arizona, and ESOs New Technology 
Telescope. 


{\sl SN~2005am:} Overall, the synthetic spectra agree with the observations including the Doppler shift of absorption features of lines, e.g. of  Si~II seen
from $\approx 6{,}080\unit{\AA}$ to $6{,}200\unit{\AA}$ at later times. 
 For the early evolution of LCs, the rise of broad features between $4{,}200\unit{\AA}$ and $5{,}000\unit{\AA}$ is noteworthy. 
At $10\unit{days}$, a significant fraction of iron-group elements are doubly-ionized with many lines of moderate opacity in the $B$-band wavelength region. 
Higher ionization shifts the strong absorption line towards the UV leaving higher excitation lines in the optical \citep{h95}.
As a result, the thermalization is higher in $B$ than $V$ in a scattering-dominated `photosphere'. 
By maximum light, the spectrum in the $B$-band is dominated by singly-ionized species, which has stronger lines in the optical.
About one week past maximum light, the spectral features in the $V$-band change from being dominated by \ion{Si}{2}, \ion{S}{2}, and \ion{Ca}{2} to those of  
\ce{Fe}-group elements, marking the end of the photospheric phase.
For example, the early \ion{Si}{2} feature is increasingly dominated by blends of singly-ionized Fe-group elements, which subsequently dominate the spectrum. Even some two months after the 
explosion, the optical emission is dominated by a quasi-continuum and broad features formed by multiple, overlapping lines.    
In our models, the strongest lines are optically thick and, consequently, are not the dominant contributor to the broad features. 
Despite the overall agreement, the agreement between synthetic and observed individual spectral features degrade at late times when forbidden lines become more important.
For example, the feature at about $5300\unit{\AA}$ is too weak some 65 days after the explosion but the continuum is increased. As a possible 
explanation, we offer that the energy deposited will be emitted partially in the optically thin wavelength region, and the energy flow
uses  as channel transitions with lowest excitation, which happen to be in the optical. As a result, the optical broadband photometry 
may be rather resilient to uncertainties in the atomic models as long as forbidden lines at NIR and MIR wavelengths remain unimportant 
for the energy balance \citep{2015ApJ...798...93T}. 

{\sl SN~2005ke:} This object is beyond the ``$\Delta m_{15}$-cliff'', which is characterized by rapid changes in the LC properties (see Fig.~\ref{observations}). Thus, we added some line identifications. Our 
Model 14 is slightly faster declining and dimmer compared to SN~2005ke. As a result, better agreement may be obtained by  
changing the parameters and/or an increased reddening correction. Moreover, the \ce{{}^{56}Ni} production in subluminous SNe~Ia 
is dominated by deflagration burning and, thus, three-dimensional effects of the burning front are expected to become important (see Sect.~\ref{DiscussionConclusions}).

The evolution of SN~2005ke is similar to SN~2005am, but here we emphasize the spectral differences.
Overall, the synthetic spectra agree with the observations including the smaller Doppler shifts compared to the normal bright model.
The \ion{Si}{2} absorption is seen between $\approx 6{,}150\unit{\AA}$ and $6{,}250\unit{\AA}$ at early and intermediate times, respectively. 
As expected from the lower luminosity of Model 14 compared to SN~2005ke, 
it shows slightly smaller photospheric velocities by about $800\unit{km}\unit{s^{-1}}$ at maximum light. 
Fe-group elements remain singly-ionized at all times even through 10 days. The theoretical spectra show persistent  
\ion{C}{2} at about $6{,}500\unit{\AA}$, and the optical \ion{Si}{2} feature remains strong beyond 10 days past maximum light. Within our series of 
$\mch$ models towards subluminous models, this can be understood in terms of an increased production of intermediate mass elements. 
 At late times, the spectral features of the model are narrower compared to SN~2005ke and often show
multiple components in the $65\unit{day}$ spectrum at e.g. $\approx 5{,}100$, $5{,}800$, and $6{,}000\unit{\AA}$. Moreover, the \ion{Co}{2} and \ion{Fe}{2} features at $\approx 
5{,}200-5{,}400\unit{\AA}$ are too weak in the synthetic spectra both at $40$ and $65\unit{days}$. The optical LCs are rather stable compared to spectra because the
cooling is dominated by optically thin lines, which mostly originate from lower excitation levels. With time, an increasing amount of 
radioactive decay energy is deposited above the photosphere by positrons. As usual, the lowest
energy levels have the highest population numbers. The UV is still optically thick. Lines in the $B$-band tend to originate from lower levels 
than in $V$, resulting in increasingly blue colors. However, a lack of cooling channels, e.g. by missing atomic data for many forbidden lines, 
causes steeper than the observed gradient in $(B-V)(t)$ for some  beyond $40-60\unit{days}$ past maximum (Figs. \ref{Series_2} \& \ref{Series_1}).
}
\end{appendix}

\bibliography{article}

\clearpage

\begin{table}
\begin{center}
\caption{Properties and stretched $\Delta m_{15,s}$ in $B$ and $V$ for all forty-three SNe~Ia of the CSP~I sample in this paper.
The first group of four SNe is used in Sect.~\ref{successes-and-limitations} for qualitative validation of the theoretical light curve and color 
relations for normal and subluminous SNe~Ia.
This sample is supplemented with seventeen of the CSP~I sample that have excellent coverage of the rise and decline, in order to 
compare the shape of the empirical and theoretical color relations in Sect.~\ref{Analysis2}.
All forty-three supernovae are used to study the brightness decline relations. 
We give the name, host galaxy, redshift $z_{\rm CMB}$ corrected for the CMB-dipole, and brightness decline rates.}
\begin{tabular}{|c||c|c|c|c|}
\hline  Object          & Host    & $z_{\rm CMB}$  & $\Delta m_{15,s}(B)$ & $\Delta m_{15,s}(V)$\\
\hline 
\hline  SN~2004eo       & NGC6928 & 0.01473   &  1.400 & 0.800  \\
\hline  SN~2005M        & NGC2930 & 0.02297   &  0.961 & 0.516 \\
\hline  SN~2005am       & NGC2811 & 0.00897   &  1.496 &0.857    \\
\hline  SN~2005ke       & NGC1371 & 0.00448   &  1.752 &1.271 \\
\hline 
\hline  SN~2004ef       & UGC12158& 0.02977   &  1.433 &0.800  \\
\hline  SN~2005el       & NGC1819 & 0.01489   &  0.969 &0.576  \\
\hline  SN~2005iq       & MGC0108 & 0.03293   &  1.220 &0.832 \\
\hline  SN~2005ki       & NGC3332 & 0.02037   &  1.242 & 0.822 \\
\hline  SN~2006D        & MGC0133 & 0.00964   &  1.395 & 0.873 \\
\hline  SN~2006X        & NGC4321 & 0.00524   &  1.164 &0.751 \\
\hline  SN~2006bh       & NGC7329 & 0.01049   &  1.431 &0.823 \\
\hline  SN~2006gt       & 2MASX J000561810-013732 & 0.04364   &  1.641 & 1.103 \\
\hline  SN~2007S        & UGC5378 & 0.01502   &  0.797 &0.499 \\
\hline  SN~2007af       & NGC5584 & 0.00629   &  1.210 &0.615 \\
\hline  SN~2007ai       &  MCG -04-38-004  & 0.03166   &   --    &0.594 \\
\hline  SN~2007le       & NGC7721 & 0.00552   &  0.935 &0.633 \\
\hline  SN~2007on       & NGC1404 & 0.00618   &  2.011 &1.110 \\
\hline  SN~2008bc       & KK1524  & 0.01571   &  0.833 &0.695 \\
\hline  SN~2008fp       & ESO 428-G14  & 0.00629    & 0.749 &0.636 \\
\hline  SN~2008hv       & NGC2765 & 0.01358   &  1.323 &0.756 \\
\hline  SN~2011iv       & NGC1404 & 0.00618   &   1.723 & 1.071 \\
\hline  
\hline SN~2004gs   & MCG +03-22-020        & 0.02750    & 1.703& 0.945    \\
\hline SN~2005A    & NGC 958        & 0.01834   &   1.242& 0.731   \\
\hline SN~2005W    & NGC 691        & 0.00795   &   1.102& 0.618               \\
\hline SN~2005al   & NGC 5304        & 0.01329   &   1.233& 0.740             \\
\hline SN~2005eq   & MCG -01-09-006        & 0.02835   &   0.910& 0.561           \\
\hline SN~2005hc   & MCG +00-06-003        & 0.04498   &    0.912& 0.609        \\
\hline SN~2005hj   & SDSS J012648.45-011417.3        & 0.05698   &    0.912& 0.609        \\
\hline SN~2005kc   & NGC 7311        & 0.01389   &   1.292& 0.741   \\
\hline SN~2005na   & UGC 3634        & 0.02681    &  1.025& 0.628 \\
\hline SN~2006et   & NGC 232        & 0.02118   &   0.993& 0.547            \\
\hline SN~2006gj   &UGC 2650         &  0.02775   &  1.745& 0.957  \\
\hline SN~2006hx   & 2MASX J01135716+0022         &  0.04440   &   1.692& 1.018         \\
\hline SN~2006kf   &  UGC 2829       &  0.02080   &    1.738& 0.881      \\
\hline SN~2007A    & NGC 105        &  0.01646  &     0.821& 0.542    \\
\hline SN~2007ax   & NGC 2577        &  0.00764  &    1.841& 1.420   \\
\hline SN~2007ba   & UGC 9798        &  0.03905  &    1.891& 1.369 \\
\hline SN~2007bc   & UGC 6332        &  0.02185  &    1.693& 0.834            \\
\hline SN~2007bd   & UGC 4455        &   0.03194  &   1.168& 0.723        \\
\hline SN~2007ca   & MCG023461        &  0.01507  &     0.940& 0.562 \\
\hline SN~2007mm   & --        &  0.06542  &     1.836& 1.487     \\
\hline SN~2008gp   & MCG +00-9-74         &  0.03282   &    1.086& 0.593   \\
\hline SN~2008ia   & ESO 125- G 006        &  0.02260  &    1.338& 0.752  \\
\hline 
\end{tabular}
\label{table_obs}
\end{center}
\end{table}

\clearpage
\begin{sidewaystable}
\begin{center}
\caption{Properties of our $\mch$ delayed-detonation models and two double-detonation models (HeD6 and HeD10). 
We give the main sequence mass $M_{\rm MS}$, metallicity $Z$, central density $\rho_c$, transition density $\rho_{\rm tr}$, and the resulting total \ce{{}^{56}Ni} production $M_{\rm Ni}$. 
For the HeD-models, we assume a $C/O=1$ core plus a He-shell with solar metallicities. 
The optical properties, decline rate $\Delta m_{15,s}$, peak magnitudes $M_{B,V}$, color and rise time at $V$-band maximum $(B-V)_{\rm max}$, and $t_s$ are given in the natural system of the Swope 
	and du Pont telescopes. In addition, we give the color-stretch $s_{BV}$, taking into account the model uncertainties in $t_{\rm max}$ and the time of maximum in $(B-V)$. For the uncertainties of the observables, see the text and the corresponding references given in the last column. 
Note that the HeD models are based on low-resolution calculations with 274 depth points.}
\begin{tabular}{|c||c|c|c|c|c|c|c||c|c|c|c|c|c|c||c|}
	\hline \multirow{2}{*}{Model} & $M_{\rm MS}$& $Z$ & $M_{\rm WD}$ &  $\rho_c$ &$ \rho_{\rm tr}$ & $M_{\rm Ni}$  & $M_V$ &$\Delta m_{15,s}(V)$ &  $M_B$ & $\Delta m_{15,s}(B)$ &  $(B-V)_{\rm max}$ &  $t_s$ & $s_{BV}$  & $\log(L_{\rm bol})$ &  Refs. \\
	       & $[\unit{\!\!M_\odot}]$& $[\unit{\!\!Z_\odot}]$ & $[\unit{\!\!M_{\odot}}]$ &  $[10^9\unit{g/cm^{3}}]$ &$ [10^6\unit{g/cm^{3}}] $ & $[\unit{\!\!mag}]$  & $[\unit{\!\!mag}]$ &$[\unit{\!\!mag/day}]$ &  $[\unit{\!\!mag}]$ & $[\unit{\!\!mag/day}] $ &  $[\unit{\!\!mag}]$ &  $[\unit{\!\!day}]$ & $ $  & $[\unit{\!\!erg/s}]$ &   \\
\hline 
\hline 08 & 5. & 1. &$\mch$ & 2. &  8. &  0.095  & -17.22  & 1.41 & -16.76  & 1.99 & 0.487 & 14.38  & 0.33-0.60 & 42.21 &  (1)     \\
\hline 08r1 & 5. & 1. & $\mch$ & 1.1  & 8.  &  0.154    &  -17.45  & 1.40 & -17.01   & 1.95 & 0.45 & 14.98  & 0.31-0.55  & 42.54 &  --    \\
\hline 08r2 & 5. & 1. &  $\mch$ & 0.5 &  8.  &  0.188   &   -17.63  & 1.39 & -17.24   & 1.93 & 0.41 & 15.26  & 0.29-0.53  & 42.70 &  --      \\
\hline 10 & 5. & 1. &$\mch$ & 2. & 10. &  0.107     & -17.35  & 1.38 & -16.94  & 1.89 & 0.427 & 14.47  & 0.40-0.60    & 42.49 &  (1)      \\
\hline 12 & 5. & 1. & $\mch$ & 2. & 12.  &  0.137       & -17.63  & 1.28 & -17.26  & 1.85 & 0.397 & 14.73  & 0.44-0.63  & 42.57 & (1)       \\
\hline 14 & 5. & 1. & $\mch$  & 2.  & 14. &   0.154    &  -17.82  & 1.24 & -17.50  & 1.81 & 0.237 & 14.86  & 0.47-0.63  & 42.69 & (1)  \\
\hline 16 & 5. & 1. & $\mch$ & 2. &  16. &  0.268    &  -18.37  & 1.18 & -18.22  & 1.67 & 0.157 & 15.37  & 0.50-0.67   & 42.84 & (1)   \\
\hline 16r1 & 5. & 1. &  $\mch$  & 1.1 & 16. & 0.324  &   -18.46  & 1.16 & -18.29   & 1.65 & 0.152 & 15.97 & 0.48-0.65 & 42.90  &  --    \\
\hline 16r2 & 5. & 1. &  $\mch$ & 0.5 &  16.  & 0.353  &    -18.57  & 1.15 & -18.43   & 1.61 & 0.15 & 16.37 & 0.46-0.64  & 42.91   &  --     \\
\hline 18 & 5. & 1. & $\mch$ & 2. &  18. &  0.365   &   -18.72  & 1.08  & -18.64  & 1.57 & 0.107 &  16.62 & 0.57-0.69  & 42.98 &  (1)   \\
\hline 18mix & 5. & 1. & $\mch$ & 2.  & 18. &  0.365 &       -18.71 & 0.8 & -18.58 &  1.37 & -0.012 &   16.62 & 0.75-0.92 &42.96  & mixed$^{+\!\!\!+}$\\
\hline 20 & 5. & 1. &  $\mch$ & 2. &  20. &  0.454   &   -18.96 &  0.89 & -18.94  & 1.47 & 0.037 & 17.35 & 0.64-0.70 & 43.06&  (1)    \\
\hline {\bf 23}& {\bf 5.} & {\bf 1.}& $\mch$ &{\bf 2.} & {\bf 23.} & {\bf 0.561}  & {\bf   -19.21} & {\bf 0.73} &{\bf -19.24}  & {\bf 1.23} &{\bf -.012}
 & { 18.24 }& {\bf  0.78-0.86 } & {\bf 43.17 } &{\bf (1)} \\
\hline 23m2 & 1.5 & 1. &$\mch$ & 2. & 23. &  0.589 &     -19.25  & 0.76 & -19.32  & 1.29 & 0.037 & 18.12  & 0.78-0.86  & 43.15 & (2)    \\
\hline 23m3 & 3. & 1. & $\mch$ & 2. & 23. &  0.567  &      -19.20  & 0.77 & -19.29  & 1.31 & -.062 & 18.19  & 0.79-0.86  &43.15 &  (2)     \\
\hline 23m4 & 7. & 1. & $\mch$ &  2. & 23. &  0.516  &      -19.11  & 0.74 & -19.18  & 1.25 & -.0425 & 19.74  & 0.84-0.90 & 43.13 &  (2) \\
\hline 23z2 & 5. &0.01& $\mch$ &   2. & 23. &  0.549 &       -19.15  & 0.73 & -19.28  & 1.23 & -.082 & 18.52  & 0.80-0.87  & 43.17   &  (2)  \\
\hline 23z3 & 5. &0.1& $\mch$ &  2. & 23. &  0.541    &  -19.14  & 0.73 & -19.25  & 1.25 & -.0625 & 18.92  & 0.81-0.87 & 43.17 & (2)   \\
\hline 23d1 &5.& 1. & $\mch$&  0.5 & 23. & 0.65 &     -19.24  &  0.72 & -19.29 &   1.23 & -.014 & 18.2 & 0.78-0.90  & 43.20  & (3,4)$^+$  \\
\hline 23d2 & 5.& 1. & $\mch$&   1.1 & 23. & 0.61 &     -19.23  &  0.72 & -19.28 &   1.23 & -.013 & 18.2 & 0.78-0.88  & 43.19  & (3,4)$^+$ \\
\hline 23d4 & 5.& 1. & $\mch$&   4. & 23. & 0.55 &     -19.19 &  0.73 & -19.21 &   1.24 & -.010 & 18.2 & 0.78-0.84  & 43.16  & (3,4)$^+$ \\
\hline 23d5 & 5.& 1. & $\mch$&   6. & 23. & 0.54 &     -19.17  &  0.73 & -19.19 &   1.24 & -.009 & 18.2 & 0.78-0.82  & 43.16  & (3,4)$^+$ \\
\hline 25 & 5. & 1. &  $\mch$ & 2. &  25. &  0.603  &    -19.29  & 0.68 & -19.32  & 1.24 & -.0125 & 18.58  & 0.92-1.02 & 43.22 &  (1)  \\
\hline 27 &  5. & 1. & $\mch$ & 2. & 27. &  0.645 &     -19.35  & 0.64 & -19.38  & 0.95 & -.003 & 18.80  & 0.99-1.08  & 43.25 & (1)   \\
\hline HeD6 & n/a  &  1. &0.6+0.22  & 0.013 & n/a& 0.252 &   -18.46  &  1.14 &  -18.36  & 1.45 & 0.09  & 14.2 & 1.0-1.6   & 42.89 &      (5)   \\
\hline HeD10 & n/a & 1. &0.8+0.16  &   0.029& n/a&  0.526  &   -19.49  &  1.03 & -19.42  & 1.35  & 0.05 & 15.0 & 0.8-1.1 & 43.35 &  (5)  \\
\hline 
\end{tabular}
\label{table_mod}
\end{center}
$(1)$ \citet{HGFS99by02}\\
$(2)$ \citet{dominguez02}\\
$(3)$ \citet{2006NewAR..50..470H}\\
$(4)$ \citet{tiara15}\\
$(5)$ \citet{hk96}\\\\
$^{+}$ { To produce very similar outer layers, the DDT has been triggered at the same pre-expansion corresponding to small
 variations in  $\rho_{\rm tr}<3\%$ within the 23d-series.} \\
$^{+\!\!\!+}$ Same as Model 18 but the abundances have been allowed to mix up to an expansion velocity of $8{,}000\unit{km}\unit{s^{-1}}$.\\
\end{sidewaystable}
\clearpage

\begin{table}
\begin{center}
\caption{Consistency checks for theoretical models and $\Delta m_{15,s}$ and CMAGIC based on fits to individual SNe~Ia (Column 1).
In Column 2, we give the theoretical $(B-V)_0$ implied by the fitting procedure (a) and actual values of the models at maximum light (b). 
In Columns 3 -- 5 we give the distance moduli based on the fits, on redshifts, and the average, minimum, and maximum values of the  
data provided by the NASA/IPAC Extragalactic Database (NED). 
The latter values are based on a wide variety of methods applied to the individual host galaxies.}
\begin{tabular}{|c|c|c|c|c|c|}
\hline  Object  &  $(B-V)_0$ (a) & $(B-V)_0$ (b) & $m-M$ & $(m-M)(z_{\rm CMB})$ & $(m-M)_{\rm ave,\,min,\,max}$ \\  
\hline 
\hline  SN~2005M   &  -0.022 & -0.030 & 35.21 & 35.21  & 35.06, 34.82, 35.21 \\
\hline  SN~2004eo  &  0.033 &   -0.018 &   34.26 & 34.23 & 33.74, 32.80, 34.66 \\
\hline  SN~2005am  & -0.089  &   -0.030 & 32.35 & 32.74 &  32.63, 31.66, 35.26 \\
\hline  SN~2005ke  &  0.530  &    0.460 &   31.64 & 31.69 & 32.09, 31.91, 32.49 \\
\hline                       
\hline  SN~2004ef  & 0.038 &    -0.018 &   35.71 & 35.70& 35.94, 34.78, 35.53 \\
\hline  SN~2005el & -0.062 &    -0.030 &   34.56  & 34.12 & 33.97, 33.29, 34.29 \\
\hline  SN~2005iq  & -0.011 &   -0.012 &   35.98 & 35.91 & 35.92, 35.26, 36.20 \\
\hline  SN~2005ki  &  0.003 &     0.010 &   34.72 & 34.67 & 34.47, 31.73, 34.91 \\
\hline  SN~2006D    & 0.127 &     0.046 &   32.93 & 32.91 & 32.73, 32.23, 33.01 \\
\hline  SN~2006X    &  0.077 &    -0.030 &   31.67 & 31.85 &  31.04, 30.20, 32.23  \\
\hline  SN~2006bh   & -0.024 &    -0.030 &   33.63 & 33.43 & 33.20, 31.60, 33.68 \\
\hline  SN~2006gt    &  0.207 &   0.084 &   36.68 & 36.51 & - \\
\hline  SN~2007S    & -0.034  &  -0.030 &   34.02 & 33.96  &33.92, 33.36, 34.22 \\
\hline  SN~2007af   &  -0.034 &   -0.030 &   32.33 & 31.94 & 31.78, 30.69, 32.67 \\
\hline  SN~2007ai   &0.126 &   -0.030 &   35.47 & 35.76 & 35.69, 35.09, 36.01 \\
\hline  SN~2007le &-0.12  &  -0.030 &   32.05 & 32.39 & 31.68, 31.19, 32.24 \\
\hline  SN~2007on   & 0.150 &    0.124 &   31.38 & 32.31 & 31.40, 30.67, 32.23 \\
\hline  SN~2008bc   & 0.046 &   -0.030 &   34.78 & 34.15 & 34.04             \\
\hline  SN~2008fp   & 0.108  &  -0.030 &  32.75 & 32.02 & 31.82 \\
\hline  SN~2008hv   &     0.063  &  -0.030 &   34.03 & 33.75 & 33.82, 33.57, 33.99 \\
\hline  SN~2011iv   &  0.068  &   0.047  &   31.23 & 32.31 & 31.40, 30.67, 32.23 \\
\hline 
\end{tabular}
\label{table_fit}
\end{center}
\end{table}

\clearpage

\begin{figure}   
\begin{center}$
\begin{array}{c}
\includegraphics[angle=360,width=0.66\textwidth]{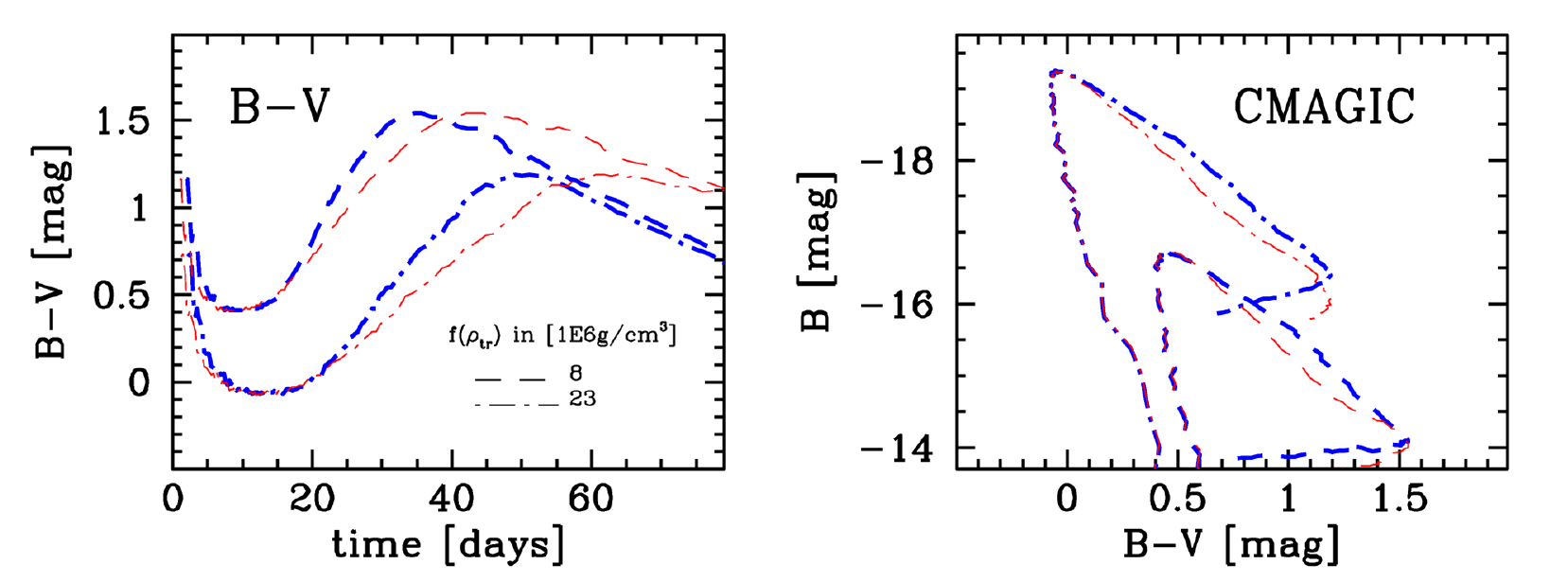}
\includegraphics[angle=360,width=0.33\textwidth]{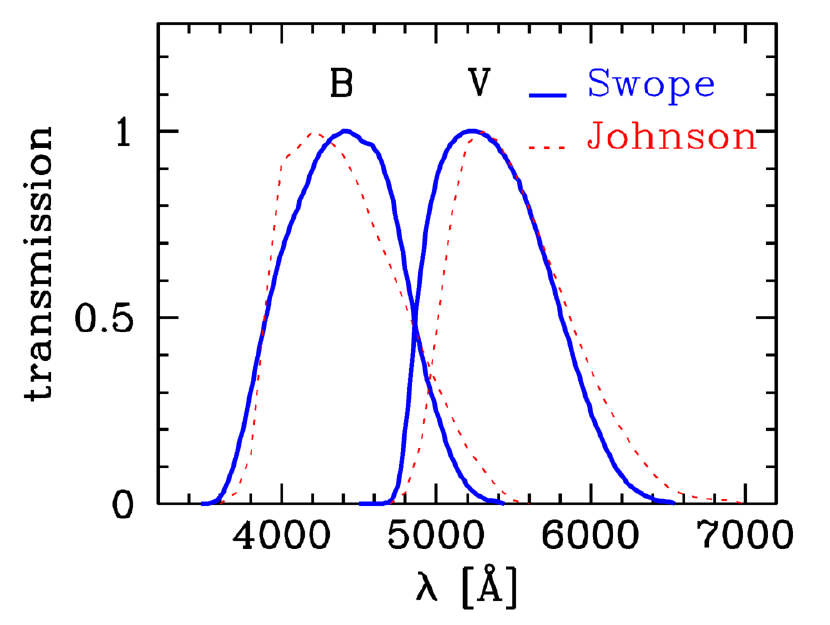}
\end{array} $
\end{center} 
\caption{
Color relations depend sensitively on the filter systems. 
As an example, we show $(B-V)(t)$ (left), CMAGIC (middle) for Models 08 (dashed) and 23 (dash-dotted) and the $B$- and $V$-band 
normalized transmission functions for the natural filter system of the Swope telescope (blue, thick solid) \citep{2011AJ....142..156S} and the  
classical Johnson (red, thin dotted) \citep{1990PASP..102.1181B} filter system, which was previously used in most of our theoretical 
studies. 
Note the importance of consistent data sets for probing SNe physics by variations in the empirical relations and for use in high-precision 
cosmology.
}
\label{filter}
\end{figure}  
 
\begin{figure} 
\begin{center}  
\includegraphics[angle=360,width=0.98\textwidth]{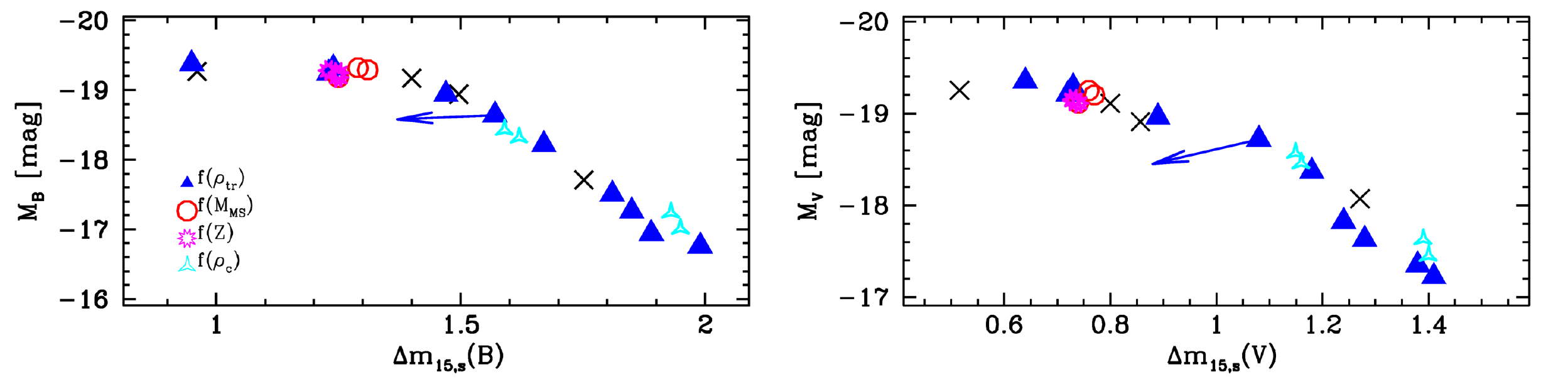}
\caption{
Same as Fig.~\ref{LC_max} but in $B$ and $V$ for the normal-bright 
SNe~2005M, 2004eo, and 2005am and the subluminous SN~2005ke
{ (ordered by decreasing absolute brightness)}.
Distances are based on the brightness decline rate $\Delta m_{15}(B)$ \citep{phillips99} rather than redshift distances.   
Note that the $\Delta m_{15}(B)$ relation can be applied for all SN~Ia with $\Delta m_{15,s}(B) \lesssim 1.7\unit{mag}$.
}
\label{obs_max}
\end{center}
\end{figure}  
 
\begin{figure}
\begin{center}   
\includegraphics[angle=360,width=0.98\textwidth]{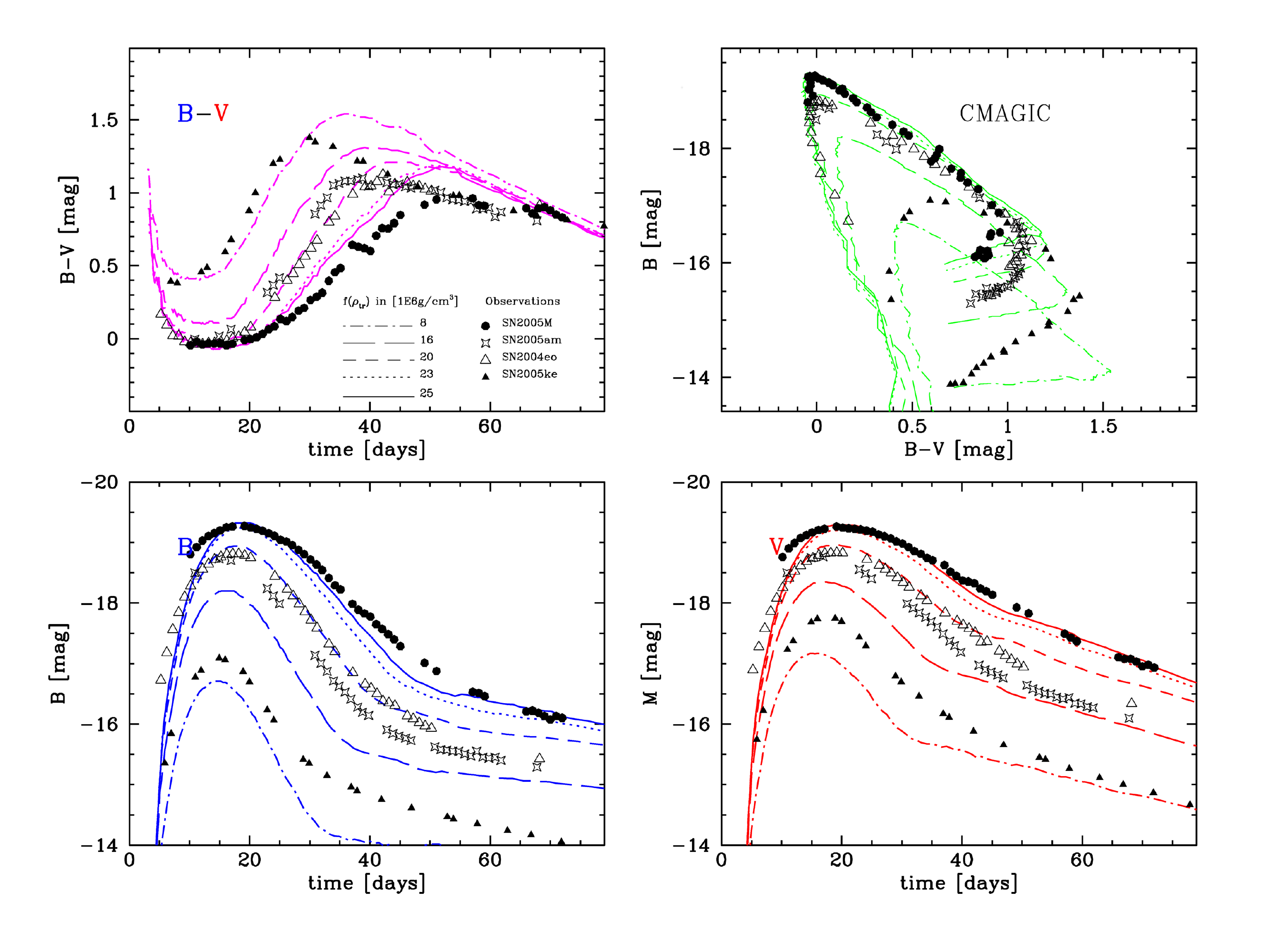}
\caption{
$(B-V)(t)$ and CMAGIC for three normal-bright SNe~Ia: 2005M, 2004eo, and 2005am, and the subluminous SN~2005ke with reddening corrections based on SNooPY \citep{contreras2010,burns11} compared with classical DDT models. 
For this comparison, we assumed rise times of $19$, $18.5$, $18.3$ and $17.2\unit{days}$, respectively, and $\Delta m_{15,s}$ based on distances (Fig.~\ref{dm15r}).
}
\label{observations}
\end{center}
\end{figure}  

\begin{figure}  
\begin{center} 
\includegraphics[angle=360,width=0.94\textwidth]{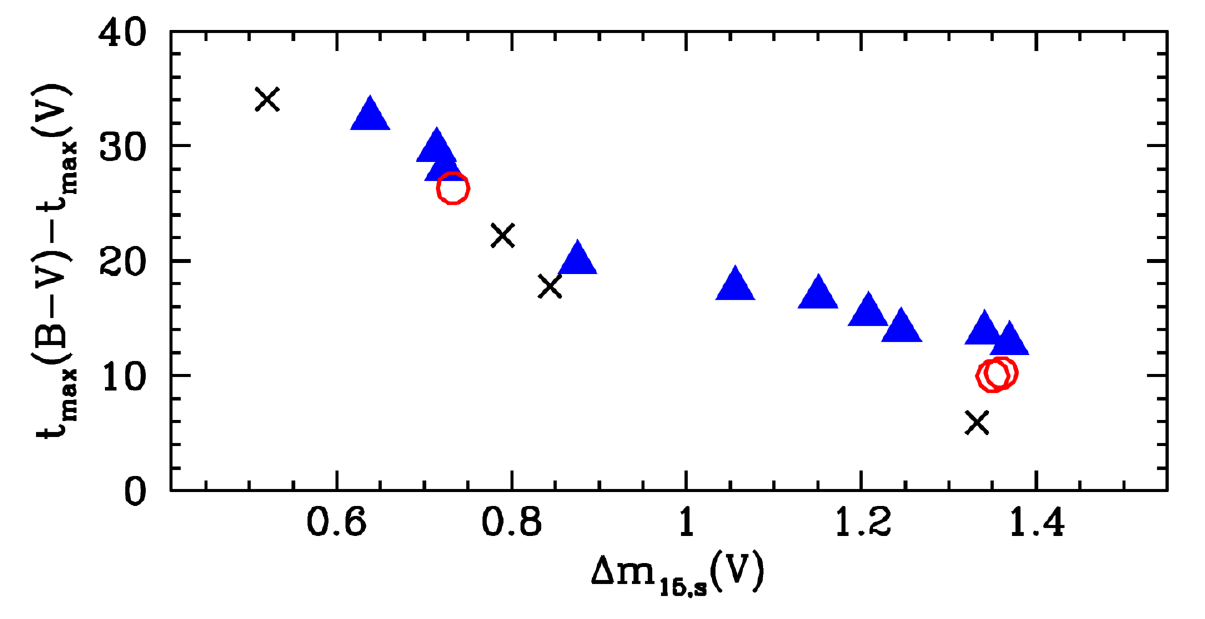}
\caption{
Difference in days between the  time of maximum in $(B-V)_0$ and the rise time to $t(V)_{\rm max}$ as a function of $\Delta m_{15,s(V)}$ 
{ for our series with variable $\rho_{\rm tr}$ (blue triangles). In addition, we give the position of low $\rho_c$ (red circles), namely our Models 
23d1, 08r1, and 08r2, and SNe~2005M, 2004eo, 2005am, and 2005ke  (ordered by decreasing absolute brightness)}. 
As discussed in the text, the uncertainties in the theoretical times are on the order of $\pm3\unit{days}$.
}
\label{burns}
\end{center}
\end{figure}

\begin{figure}   
\begin{center}
\includegraphics[angle=360,width=0.94\textwidth]{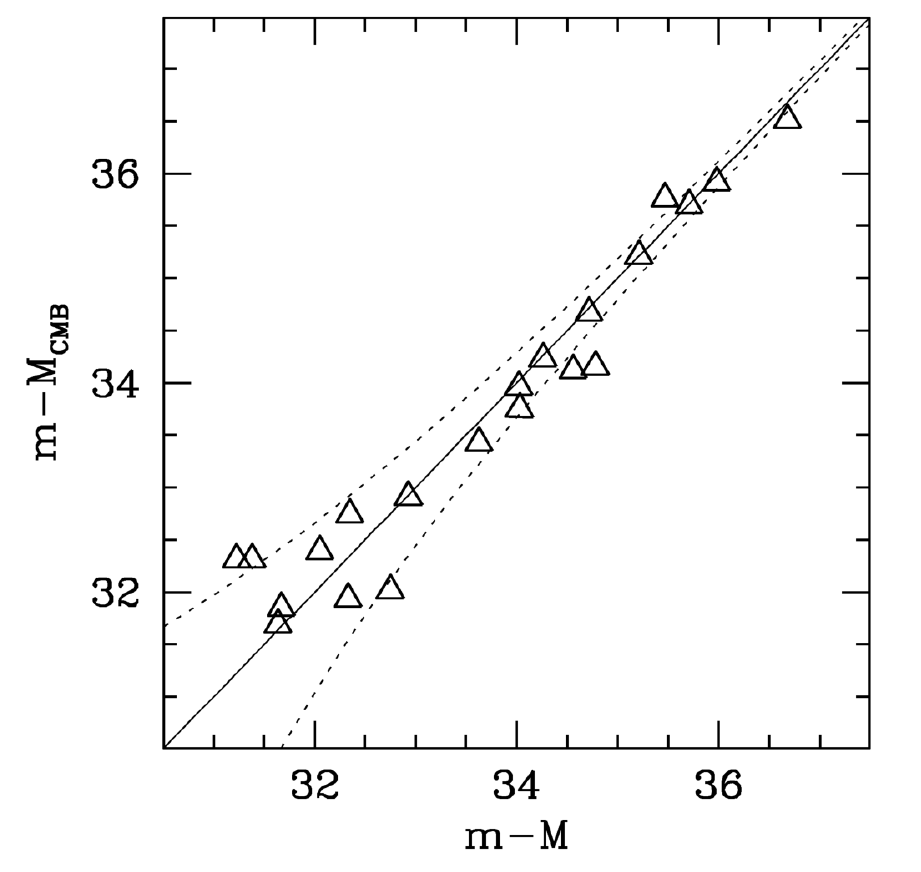}
\caption{
Comparison of the redshift-based, CMB-corrected distances and those obtained using our CMAGIC method combined with 
$\Delta m_{15,s}(B,V)$-based optimized values for the time of the explosion and reddening. 
In addition, we show the uncertainties in redshift distances due to peculiar motions, which prevent the use of redshift distances to study 
the shape of color and color-magnitude relations for many well-observed nearby objects.
}
\label{dist} 
\end{center}
\end{figure}  

\begin{figure}   
\begin{center}
\includegraphics[angle=360,width=0.94\textwidth]{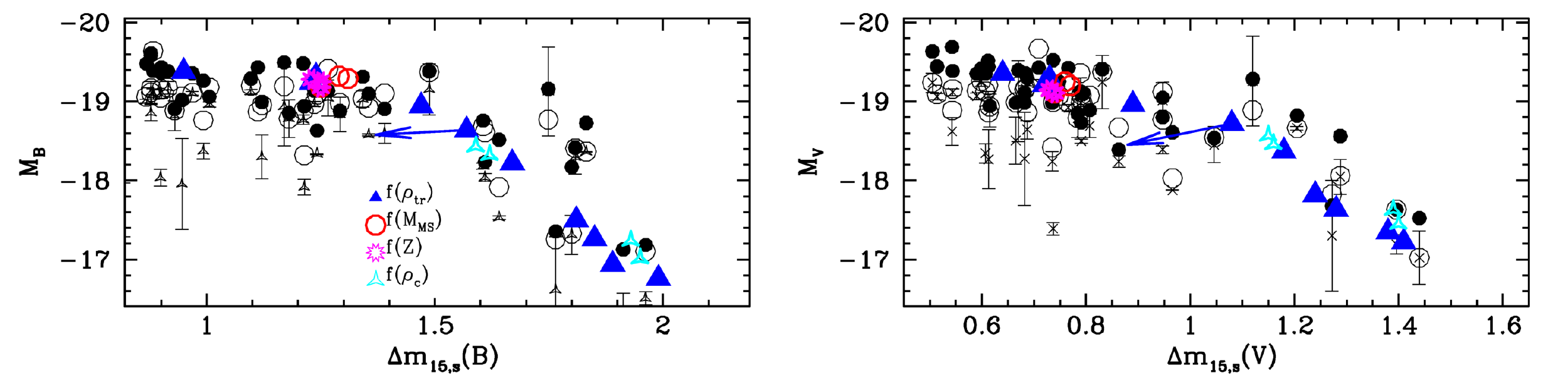}
\caption{
For illustration and as a consistency check for the models, we show $\Delta m_{15,s}(V)$ and $\Delta m_{15,s}$ optimized using  
CMAGIC for the time of the explosion and reddening (open circles) and, as a benchmark, loci of the empirical reddening corrections 
(black) based on the method of \citet{burns11}. 
In addition, we give the SN~Ia loci without reddening correction (open triangle).
For individual SNe~Ia, the related points are shifted along the y-axis. 
The importance of reddening corrections are obvious for high-precision cosmology. 
We note that our combined $\Delta m_{15}$-CMAGIC optimized fits are consistent with the semi-empirical method.
}
\label{dm15r}
\end{center}
\end{figure}  
 
\begin{figure}   
\begin{center}
\includegraphics[angle=360,width=0.98\textwidth]{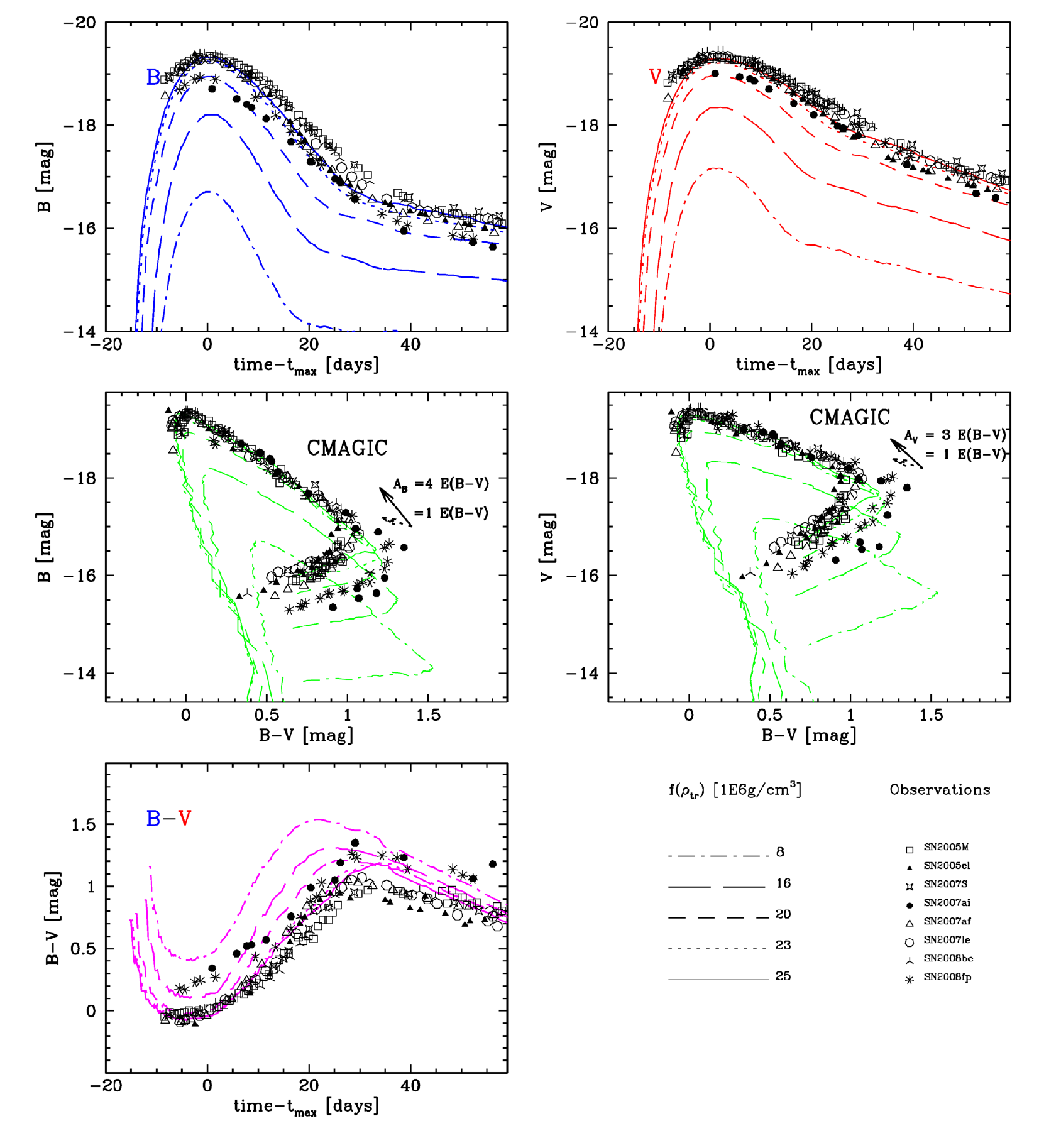}
\caption{
Comparison between four well-observed SNe~Ia with a narrow range of  ${ 0.5 < \Delta m_{15,s}(V) < 0.7\unit{mag}}$ and theoretical 
models using the same symbols as in Fig.~\ref{LC_max}. 
We compare the $B$- and $V$-band LCs (upper panels), $B(B-V)$ and $V(B-V)$ (middle panels), and $(B-V)(t)$ (lower panel). 
The observations have been shifted along the $y$- and $x$-axes employing the $\Delta m_{15,s}$ and CMAGIC-based procedure, 
distances, and standard reddening laws (see Sect.~\ref{Reddening}). 
Note that the shifts of the individual relations are not independent but based on the same set of three common parameters allowing us to 
	test for the consistency between observations and models. { In addition, the solid and dotted arrows indicated the direction of shifts by the ISM for regular and flat reddening laws, respectively.} 
For normal-bright SNe~Ia, both the observations and models show a small dispersion.
}
\label{Series_2}
\end{center}
\end{figure}  
 
 
\begin{figure}   
\begin{center}
\includegraphics[angle=360,width=0.98\textwidth]{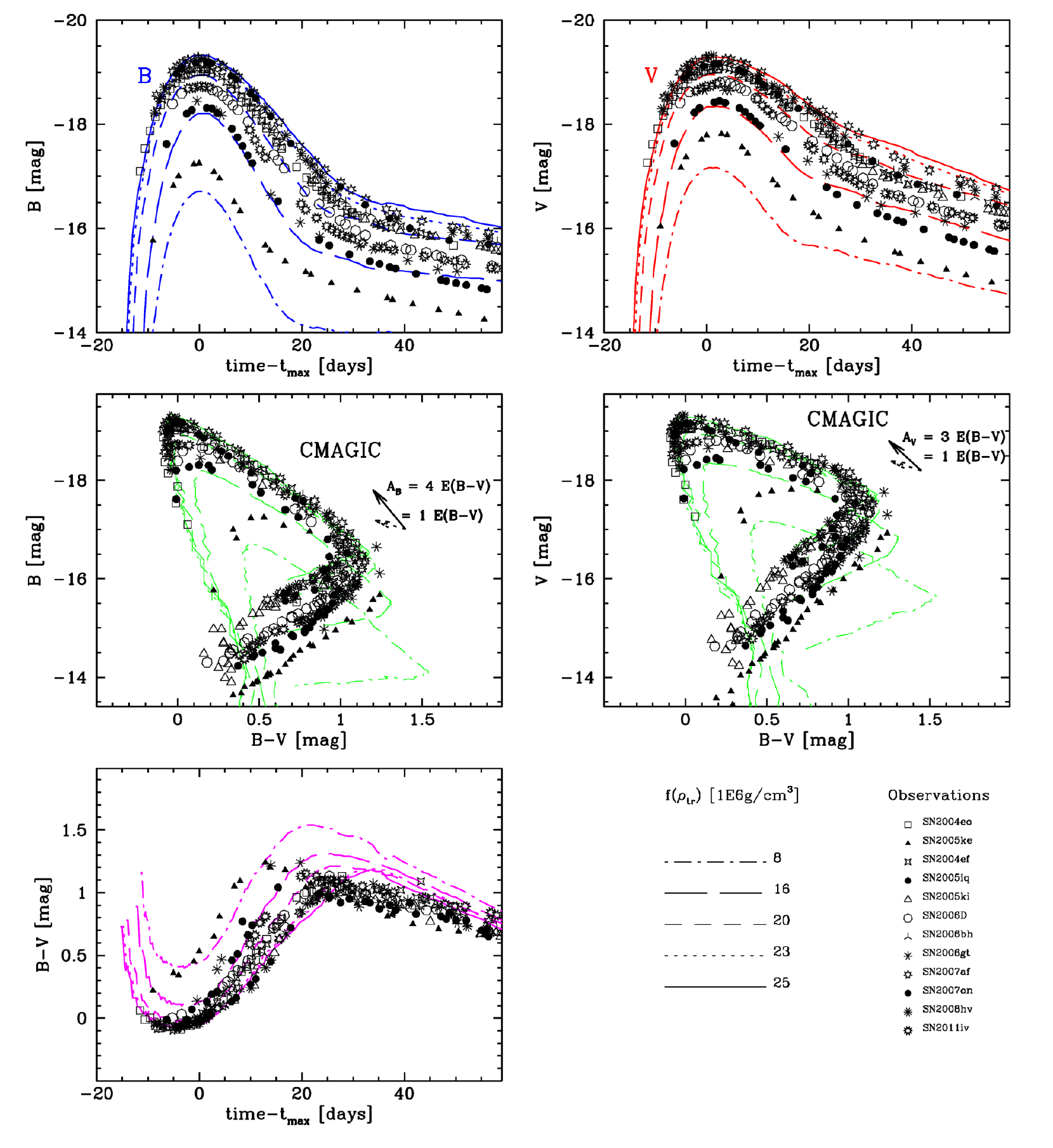}
\caption{
Same as Fig.~\ref{Series_2} but for a wider range of  ${ 0.518 < \Delta m_{15}(V) < 1.29\unit{mag}}$.
The predicted and observed shifts of the CMAGIC are consistent. 
}
\label{Series_1}
\end{center}
\end{figure}  
 
\begin{figure}   
\begin{center}
\includegraphics[angle=360,width=0.98\textwidth]{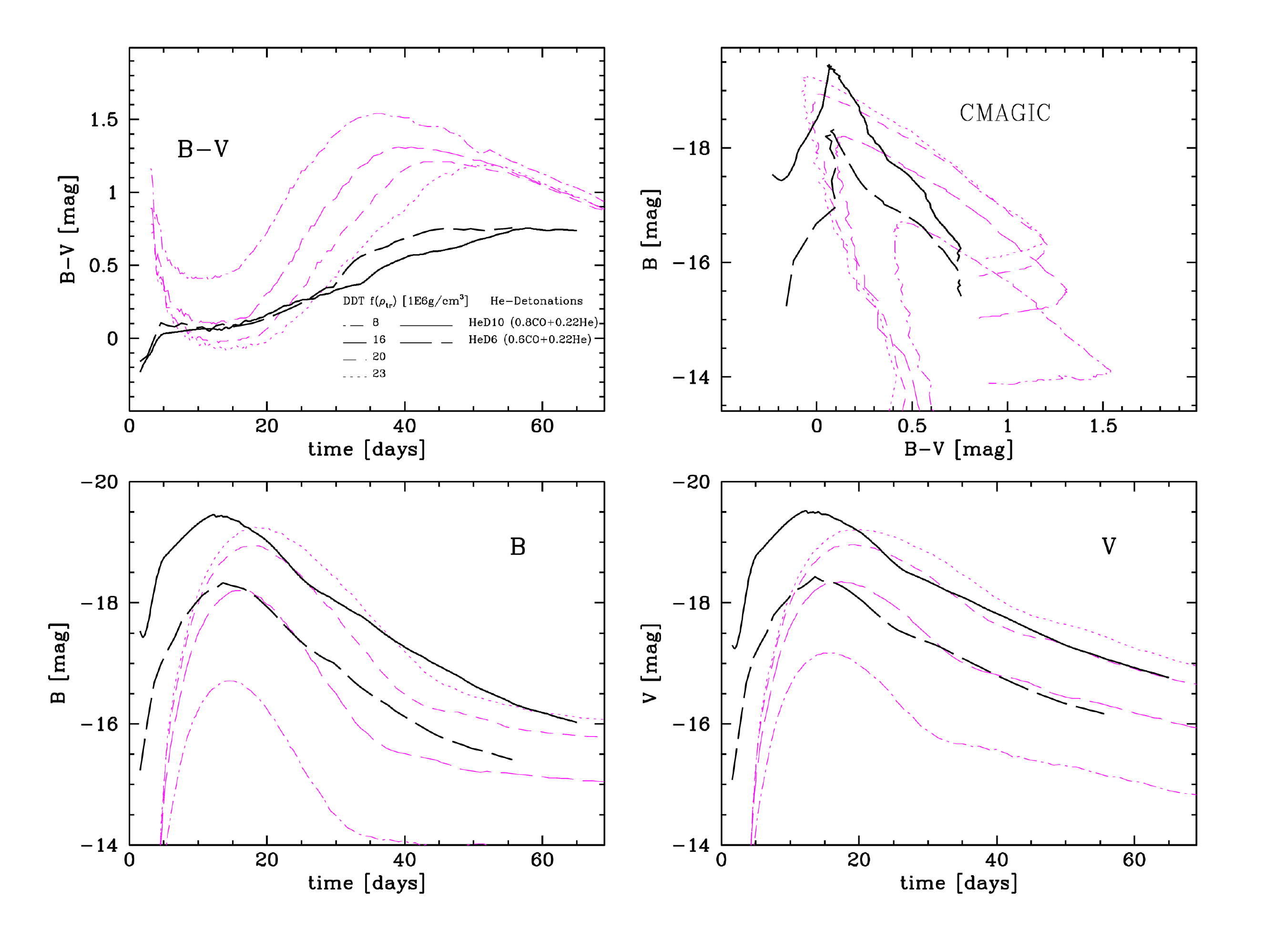}
\caption{
Color evolution $(B-V)(t)$ of double-detonation or \ce{He} detonation models (black) in comparison to our sequence of the DDT models 
shown in Fig.~\ref{LC_properties}. 
The Models HeD10 and HeD6 are shown, which consist of \ce{C}-\ce{O} WDs of $0.8$ and $0.6\unit{M_{\odot}}$ surrounded by 
$0.2\unit{M_{\odot}}$ of helium. 
They produce about $0.54$ and $0.3\unit{M_{\odot}}$ of \ce{{}^{56}Ni} corresponding to a normal-bright and subluminous SN~Ia, 
respectively \citep{hk96}.  
Compared to the normal-bright SN~Ia, low masses in Model HeD result in faster rising LCs (lower plots). 
}
\label{He_properties}
\end{center}
\end{figure}  

\begin{figure}   
\begin{center}
\includegraphics[angle=360,width=0.8\textwidth]{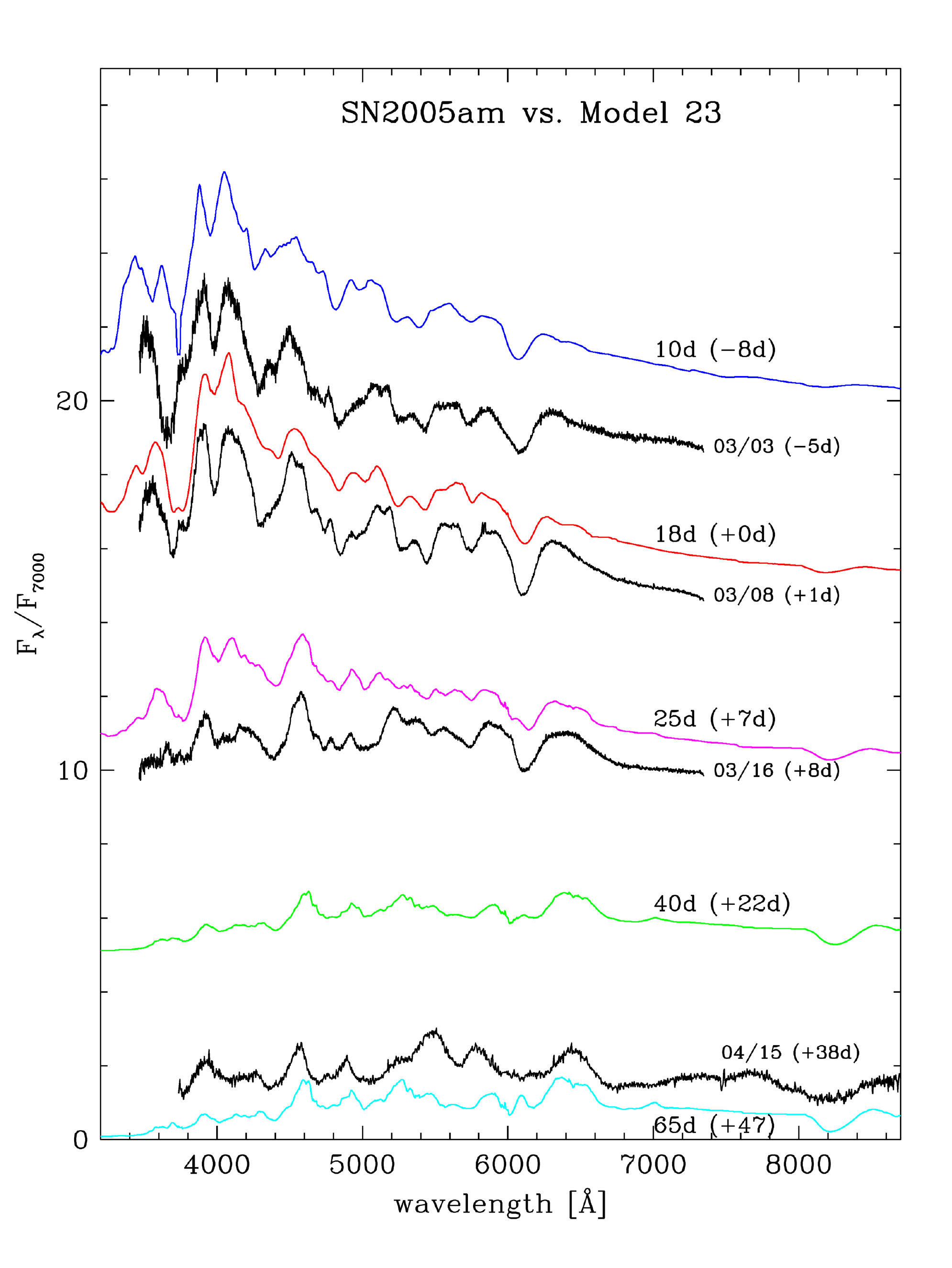}
\caption{For illustration, time series are given for synthetic and observed spectra of Model 23 and SN~2005am, respectively.   
Model 23 has been chosen because $\Delta m_{15}$ is the closest to SN~2005am (Fig. \ref{observations}).
The theoretical spectra are reddened by $E(B-V)=0.09\unit{mag}$ \citep{burns11} and redshifted (see Table~1).
The synthetic and observed spectra are labeled with respect to the time since explosion and the UT date, respectively.
  In addition, we give the time relative to theoretical and observed maximum light in brackets.
All spectra are normalized to $7{,}000\unit{\AA}$ and shown with a five unit offset for the theoretical models.
 The observations have been offset by 1.0, 9.3, 14.0 and 18.0 units to place them close to their location in time.
About one week after maximum, the spectral features in $V$ change from being dominated by Si and S to those of blends of iron group elements. Even 2 months after the explosion, the optical emission is dominated by a quasi-continuum
formed by overlapping lines. In our models, the relative rise of broad features between $4{,}200$ and $5{,}000\unit{\AA}$, the $B$-band,
is a result of the recombination of doubly- to singly-ionized  iron-group elements at the photosphere \citep{h95}.
A brief discussion and some line identifications are referred to in the Appendix.
}
\label{SN2005am}
\end{center}
\end{figure}  

\begin{figure}   
\begin{center}
\includegraphics[angle=360,width=0.8\textwidth]{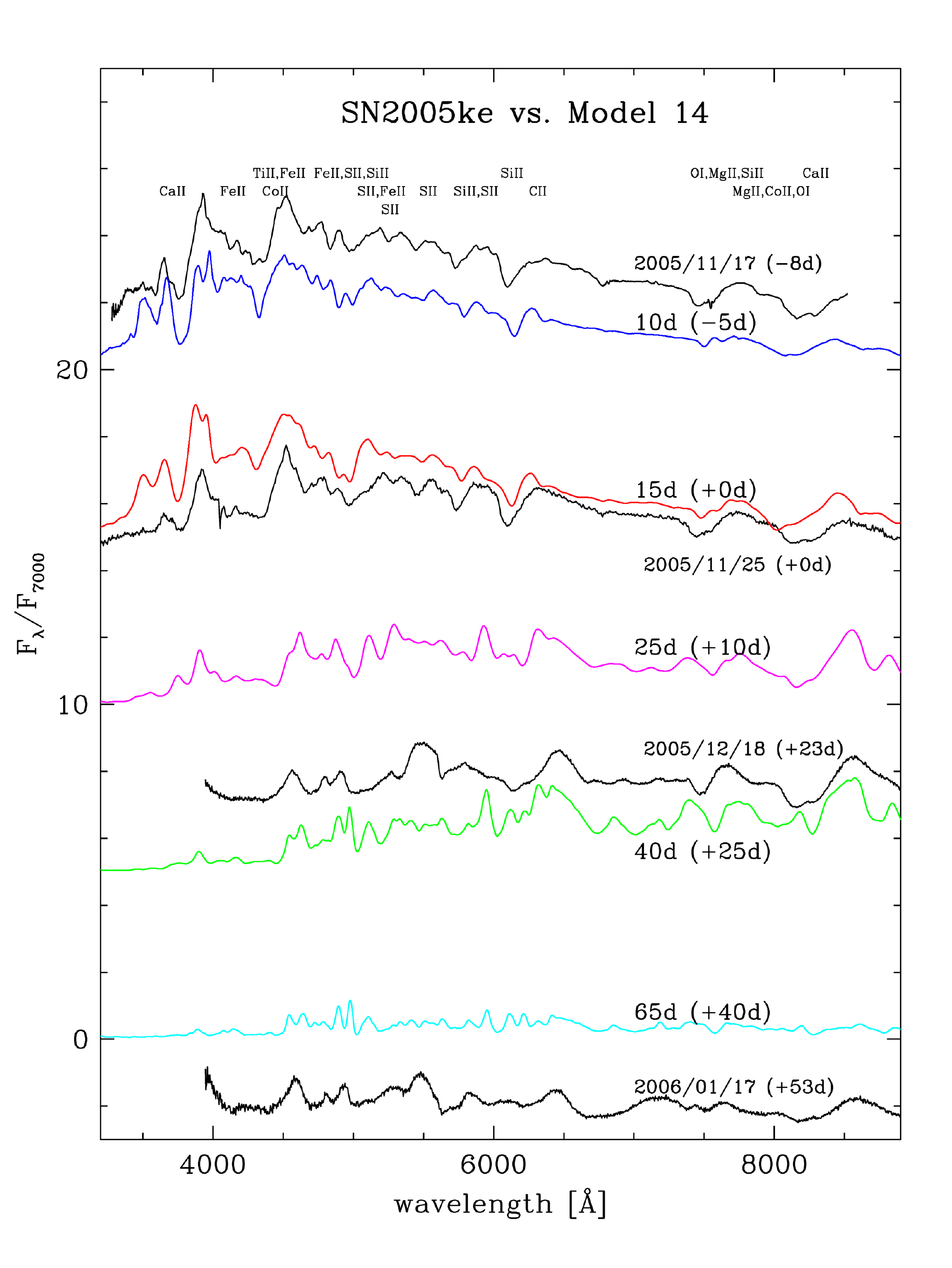}
\caption{For illustration, time series are given for theoretical and observed spectra of the subluminous Model 14 and SN~2005ke, respectively.   
Model 14 has been chosen because $\Delta m_{15}$ is the closest to SN~2005ke (Fig.~\ref{observations}).
The theoretical spectra are reddened by $E(B-V)=0.18\unit{mag}$ \citep{burns11} and redshifted (see Table 1).
The synthetic and observed spectra are labeled by the time since explosion and the UT date, respectively. In addition, we give the time relative to maximum light.
All spectra are normalized to $7{,}000\unit{\AA}$ and shown with a five unit offset for the theoretical models.
 The observations have been offset by  $-1.5$, 6.8, 14.6 and 21 units to place them close to their location in time.  
In addition, we identify some of the major contributors to spectra features at early times. By 25 days the spectra are dominated
by blends of \ion{Fe}{2} and \ion{Co}{2}. The \ion{Ca}{2} resonance features persist because \ion{Ca}{2} is produced during incomplete Si burning (see Fig.~\ref{models}). 
Beyond the $\Delta m_{15}$-cliff, the properties depend sensitively on the heating by \ce{{}^{56}Ni}. As a result, the 
photosphere recedes slightly faster in the envelope and the Doppler shifts are somewhat smaller than observed.
}
\label{SN2005ke}
\end{center}
\end{figure}

\end{document}

